\documentclass[aps,reprint,prd,preprintnumbers,superscriptaddress,nofootinbib]{revtex4-2}

\usepackage{amsmath, amssymb, amsthm}
\usepackage{mathtools}
\usepackage{mathrsfs}
\usepackage{bm}
\usepackage{slashed}   

\usepackage{graphicx}
\usepackage{multirow}
\usepackage{tikz}
\usepackage[caption=false]{subfig}
\usepackage{relsize}	
\usepackage{array}
\usepackage{float}
\usepackage{color}
\usepackage{xcolor}
\usepackage{soul}

\usepackage{siunitx}
\usepackage{hyperref}
\hypersetup{colorlinks=true,linkcolor=blue,anchorcolor=blue,citecolor=blue,
	urlcolor=black}

\usepackage{booktabs}
\usepackage{multirow}

\begin{document}
	
	\title{Heavy mesons with dynamical gluon on the light front}
	
	\author{Jiatong Wu}
	\affiliation{Institute of Modern Physics, Chinese Academy of Sciences, Lanzhou 730000, China}
	\affiliation{School of Nuclear Science and Technology, University of Chinese Academy of Sciences, Beijing 100049, China}
	\affiliation{CAS Key Laboratory of High Precision Nuclear Spectroscopy, Institute of Modern Physics, Chinese Academy of Sciences, Lanzhou 730000, China}
	
	\author{Hengfei Zhao}
	\affiliation{Institute of Modern Physics, Chinese Academy of Sciences, Lanzhou 730000, China}
	\affiliation{School of Nuclear Science and Technology, University of Chinese Academy of Sciences, Beijing 100049, China}
	\affiliation{CAS Key Laboratory of High Precision Nuclear Spectroscopy, Institute of Modern Physics, Chinese Academy of Sciences, Lanzhou 730000, China}
	
	\author{Kaiyu Fu}
	\affiliation{Institute of Modern Physics, Chinese Academy of Sciences, Lanzhou 730000, China}
	\affiliation{School of Nuclear Science and Technology, University of Chinese Academy of Sciences, Beijing 100049, China}
	\affiliation{CAS Key Laboratory of High Precision Nuclear Spectroscopy, Institute of Modern Physics, Chinese Academy of Sciences, Lanzhou 730000, China}
	
	\author{Zhi Hu}
	\affiliation{Institute of Modern Physics, Chinese Academy of Sciences, Lanzhou 730000, China}
	\affiliation{School of Nuclear Science and Technology, University of Chinese Academy of Sciences, Beijing 100049, China}
	\affiliation{CAS Key Laboratory of High Precision Nuclear Spectroscopy, Institute of Modern Physics, Chinese Academy of Sciences, Lanzhou 730000, China}

	\author{Xingbo Zhao}
	\affiliation{Institute of Modern Physics, Chinese Academy of Sciences, Lanzhou 730000, China}
	\affiliation{School of Nuclear Science and Technology, University of Chinese Academy of Sciences, Beijing 100049, China}
	\affiliation{CAS Key Laboratory of High Precision Nuclear Spectroscopy, Institute of Modern Physics, Chinese Academy of Sciences, Lanzhou 730000, China}
	
	\author{James P. Vary}
	\affiliation{Department of Physics and Astronomy, Iowa State University, Ames, Iowa 50011, USA}
	
	\collaboration{BLFQ Collaboration}
	
	\date{\today}
	
	\begin{abstract}
		We investigate the structure of charmonium, bottomonium, and $\rm B_c$ meson systems within the Basis Light-Front Quantization (BLFQ) approach, including both the quark-antiquark ($|q\bar{q}\rangle$) and quark-antiquark-gluon ($|q\bar{q}g\rangle$) Fock sectors. Our input light-front Hamiltonian incorporates a confining potential inspired by light-front holography, as well as the quark-gluon interaction from Quantum Chromodynamics. By adjusting model parameters to reproduce the mass spectra for low-lying states, we obtain the light-front wave functions for the heavy meson states. Based on these wave functions, we calculate electromagnetic form factors, decay constants, parton distribution amplitudes (PDAs), and parton distribution functions (PDFs) of the quarks and gluons in the heavy mesons. Our results for the charge radii and decay constants reasonably agree with experimental data and other theoretical approaches. The PDAs are consistent with the predictions from the earlier BLFQ calculations with an effective one-gluon exchange interaction. Furthermore, we present the first predictions within the BLFQ framework for the gluon PDFs in heavy mesons based on the light-front wave function in the $|q\bar{q}g\rangle$ sector.
	\end{abstract}
	
	\maketitle
	
	\section{INTRODUCTION}
	Quantum Chromodynamics (QCD) is the fundamental theory describing the strong interaction dynamics of hadrons. However, it remains challenging to calculate the hadron structure from QCD first principles. Nonperturbative approaches such as the lattice gauge theory~\cite{Joo:2019byq,MILC:2009mpl,Briceno:2017max,Ma:2017pxb,HAGLER201049,PhysRevD.81.034508} and the Dyson-Schwinger Equation (DSE)~\cite{Maris:2006ea,doi:10.1142/S0218301303001326,ROBERTS1994477,Bashir_2012,ALKOFER2001281} have made considerable progress in the calculations of the mass spectra and parton distributions of hadrons. As a complementary approach, the light-front Hamiltonian formalism offers direct access to the light-front distributions of partons inside hadrons. These distributions provide the three-dimensional imaging of hadrons and can be experimentally accessed through high-energy probes~\cite{BRODSKY1998299}. 
	
	Within the light-front Hamiltonian framework, Basis Light-front Quantization (BLFQ) has been developed as a nonperturbative approach to relativistic multi-body bound state problems~\cite{j.vary}. This method has been successfully applied to various Quantum Electrodynamics (QED) and QCD systems, such as the positronium~\cite{PhysRevD.91.105009,pos5}, light mesons~\cite{doi:10.1142/9789811219313_0099,s.jia,j.lan1,Lan:2019rba,Zhu:2023lst,Lan:2025fia}, heavy mesons~\cite{y.li1,s.tang,PhysRevD.102.014020,PhysRevD.105.L071901}, and baryons~\cite{PhysRevD.102.016008,PhysRevD.104.094036,Zhang:2023xfe,Zhu:2023nhl,Zhu:2024awq}. For a recent review on the applications of BLFQ to the nucleon system, see Ref.~\cite{Vary:2025yqo}.
	
	Heavy mesons formed by $c$ and $b$ quarks have long been considered as a theoretical laboratory for investigating the interplay between the nonperturbative and perturbative dynamics of QCD~\cite{Brambilla2011}. The presence of the heavy quark mass scale, which is significantly larger than the intrinsic QCD scale ($m_q\gg\Lambda_{\rm QCD}$), makes these systems amenable to perturbative analyses. On the other hand, their binding energy on the scale of $\Lambda_{\rm QCD}$ suggests that their structure is still influenced by the nonperturbative physics at the scale of $\Lambda_{\rm QCD}$. Therefore, studying heavy meson properties serves not only as a test ground of the BLFQ approach but also as an avenue for understanding the interplay between the nonperturbative and perturbative physics in QCD.
	In previous BLFQ studies, Li and Tang~\cite{y.li1,s.tang} investigated heavy meson systems within the leading $|q\bar{q}\rangle$ Fock sector, employing a light-front holographic QCD confinement supplemented by an effective one-gluon exchange interaction (OGE). This work reproduced the mass spectra of low-lying heavy meson states and the decay constants for quarkonium states, and thereby demonstrated the feasibility of the light-front approach for the heavy meson systems.

	Despite the fact that the gluon radiation in heavy mesons is suppressed compared to light mesons, gluons exist in heavy mesons as the mediators of the strong interaction between the quark and antiquark.
	The distribution of the gluons in heavy mesons thus reveals key information on the interaction between the heavy quarks.
	In the present work, we extend the study of the heavy mesons to the basis space containing one dynamical gluon, that is, $|q\bar{q}g\rangle$. By explicitly including the dynamical gluon degrees of freedom, our results serve as a more realistic initial condition for the QCD evolution, which leads to more accurate predictions of the parton structure of heavy mesons at evolved scales.
	Similar approaches have been applied to the light mesons~\cite{Lan:2025fia} and the nucleon~\cite{PhysRevD.108.094002}. By considering the distribution of the dynamical gluons in the light meson, heavy meson, and nucleon systems, we expect to obtain a systematic overall picture on the role played by the gluon in the hadrons.
	
	This paper is organized as follows. In Section~\ref{sec2}, we introduce the formalism of the BLFQ approach, including the input Hamiltonian, the basis construction, as well as the regularization and renormalization scheme. In Section~\ref{sec3}, we present the numerical results for the mass spectrum, light-front wave functions, electromagnetic form factors, charge radii, decay constants, parton distribution amplitudes, and parton distribution functions of the heavy meson systems. We summarize the paper in Section~\ref{sec4}. 
	
	\section{HAMILTONIAN FORMALISM AND THE BASIS FUNCTION REPRESENTATION}\label{sec2}
	In BLFQ~\cite{j.vary}, one diagonalizes the following mass eigenequation to obtain the invariant mass and light-front wave functions for hadrons,
	\begin{equation}
		P^\mu P_\mu |\Psi\rangle =M^2 |\Psi\rangle,
		\label{fc}
	\end{equation}
	where $P^\mu=(P^+,P^-,\vec{P}_\perp)$ is the energy-momentum operator in the light-front coordinates with $P^\mu P_\mu=P^-P^+ - \vec{P}^2_\perp$. $P^-,P^+,\vec{P}_\perp$ are the light-front Hamiltonian, longitudinal and transverse momentum operators, respectively. $M$ is the invariant mass of the hadron eigenstate $|\Psi\rangle$.
	
	In this work, we apply BLFQ to study the heavy meson states in the $|q\bar{q}\rangle$ and $|q\bar{q}g\rangle$ Fock sectors. Their light-front wave functions at the fixed light-front time ($x^+=x^0+x^3$) can be written as,
	\begin{equation}
		|\Psi\rangle=\Psi_{q\bar{q}}|q\bar{q}\rangle+\Psi_{q\bar{q}g}|q\bar{q}g\rangle,
	\end{equation}
	where $\Psi_{q\bar{q}}$ and $\Psi_{q\bar{q}g}$ denote the light-front wave functions (LFWFs) in the $|q\bar{q}\rangle$ and $|q\bar{q}g\rangle$ Fock sectors, respectively. These LFWFs encode the structural information of the heavy meson systems.
	
	In this truncated Fock space, we adopt the following input light-front Hamiltonian~\cite{Lan:2021wok}, $P^-=P^-_{\rm QCD}+P^-_{\rm C}$, where $P^-_{\rm QCD}$ contains the kinetic energy terms and QCD interactions relevant for the $|q\bar{q}\rangle$ and $|q\bar{q}g\rangle$ Fock sectors and $P_{\rm C}^-$ models the confining potential in the longitudinal and transverse directions.
	The light-front QCD Hamiltonian with one dynamical gluon in the light-front gauge $A^+=0$ is,
	
	\begin{equation}
		\begin{aligned}
			P^-_{\rm QCD}&= \int \mathrm{d}^2 x^{\perp} \mathrm{d}x^- \left\{ \frac{1}{2} \bar{\Psi} \gamma^+ \frac{m_{q0}^2+(i\partial^{\perp})^2}{i\partial^+} \Psi \right.\\
			&+\frac{1}{2} A^\mu_a [m_g^2+(i\partial^{\perp})^2] A^a_\mu + g \bar{\Psi} \gamma_{\mu} T^a A^{\mu}_a \Psi \\
			&+\frac{1}{2} \left.g^2 \bar{\Psi} \gamma^+ T^a \Psi \frac{1}{(i\partial^+)^2} \bar{\Psi} \gamma^+ T^a \Psi \right\},
		\end{aligned} 
	\end{equation}
	where $\Psi$ and $A^\mu$ represent the quark and gluon fields, respectively.
	The first two terms of $P^-_{\rm QCD}$ are the kinetic energy of the quark and gluon fields, respectively. The last two terms are the quark-gluon vertex interaction and the instantaneous gluon exchange interaction, respectively.
	Following the ``gauge principle"~\cite{Tang:1991rc,Lan:2021wok}, the instantaneous gluon exchange interaction is only applied to the $|q\bar{q}\rangle$ sector. $m_{q0}$ denotes the bare mass of the quark and $m_q$ is the physical mass. In order to compensate for the quark self-energy correction from the higher Fock sector with one dynamical gluon, we introduce a mass counterterm, $\delta m_q=m_{q0}-m_q$, for the quark in the leading Fock sector following the procedure of the sector-dependent renormalization~\cite{PhysRevD.77.085028,PhysRevD.86.085006}. An explicit gluon mass term $m_g$ is introduced as an effective parameter to regularize infrared divergences and mimic confinement~\cite{Lan:2021wok}. $T^a$ is the half Gell-Mann matrix, $T^a=\frac{\lambda^a}{2}$. $g$ is the strong gauge coupling, related to the strong coupling constant $\alpha_s$ by $\alpha_s=\frac{g^2}{4\pi}$.
	
	In addition, we generalize the holographic confining interaction in the transverse directions by introducing the longitudinal confining potential in the leading Fock sector to reproduce the three-dimensional confinement in the nonrelativistic limit~\cite{y.li1}. Accordingly, the confinement interaction $P^-_{\rm C}$ reads~\cite{y.li4,PhysRevD.104.094036},
	\begin{equation}
		P^-_{\rm C}P^+=\kappa_{T}^4 \vec{\xi}_{\perp}^2-\frac{\kappa^4_{L}}{(m_q+m_{\bar{q}})^2}\partial_x(x(1-x)\partial_x),
	\end{equation}
	where $x$ is the longitudinal momentum fraction of the quark. $\kappa_{L}$ and $\kappa_{T}$ are the strength parameters for the longitudinal and transverse confining potentials, respectively. $\vec{\xi}_{\perp} \equiv \sqrt{x(1-x)} \vec{r}_\perp$ is the holographic variable, where $\vec{r}_\perp= \vec{r}_{\perp q}-\vec{r}_{\perp \bar{q}}$, and $\partial_x \equiv(\partial / \partial x)_{\xi_{\perp}}$~\cite{BRODSKY20151}.
	
	Since in BLFQ we adopt the single-particle basis, cf.~Sec.~\ref{sec2A}, the excitation of the center-of-mass motion is allowed in the mass spectrum. In order to separate the center-of-mass motion from the intrinsic motion, we introduce a constraint term to our light-front Hamiltonian $P^-$, and our complete Hamiltonian for diagonalization reads~\cite{PhysRevD.91.105009,Vary:2025yqo},
	
	\begin{equation}
		P^{\prime-} P^{+}=P^{-} P^{+} - \big(\sum_{i} \vec{p}_{\perp i}\big)^{2} +\lambda\left(P_{\mathrm{cm}}^{-} P^{+}-2 b^{2} I\right).
	\end{equation}
	Here, the center-of-mass Hamiltonian $P^{-}_\mathrm{cm}$ reads:
	\begin{equation}
		P_{\mathrm{cm}}^{-} P^{+}=\vec{P}_{\perp}^{2}+b^{4} \vec{R}_{\perp}^{2}=\big(\sum_{i} \vec{p}_{\perp i}\big)^{2}+b^{4}\big(\sum_{i} x_{i} \vec{r}_{\perp i}\big)^{2},
	\end{equation}
	where $b$ is the harmonic oscillator basis scale parameter (see below).
	A sufficiently large Lagrange multiplier $\lambda$ will shift the states with excited center-of-mass motion to the higher part of the mass spectrum and ensure that low-lying states are only intrinsically excited.
	
	\subsection{The Basis Function Representation}\label{sec2A}
	Within the BLFQ formalism, the LFWFs and the mass spectra are obtained by solving the mass eigenequation in Eq.~\eqref{fc}~\cite{j.vary,HILLER201675}.
	This equation is solved in a basis function approach. For each Fock particle, we adopt the plane-wave basis in the longitudinal direction and the two-dimensional harmonic oscillator (2D-HO) wave functions as the basis state in the transverse directions~\cite{j.vary,PhysRevD.91.105009}.
	In the longitudinal direction, the plane-wave basis is confined within a one-dimensional box of length $2L$, so the longitudinal momentum is discretized as $p^+_i=2\pi k_i/L$, where $k_i=\frac{1}{2}$, $\frac{3}{2}$, $\frac{5}{2}\dots$ for quarks and $k_i= 1,2,3\dots$ for gluons. We neglect the zero mode for gluons. The longitudinal basis function is written as $e^{-\mathrm{i} p^+_i x^-/2}$ in the longitudinal coordinate $x^-$, and the longitudinal momentum fraction is defined as $x_i=p^+_i/P^+=k_i/K$ for the $i$-th parton~\cite{PhysRevD.91.105009}.
	In momentum space, the orthonormalized 2D-HO wave functions are given by
	\begin{equation}
		\begin{aligned}
			\phi_{n m}&\left(b'; \vec{p}_{\perp}\right)=\frac{1}{b'} \sqrt{\frac{4 \pi n !}{(n+|m|) !}}\left(\frac{p_{\perp}}{b'}\right)^{|m|} \\
			&\times \exp \left(\frac{-p_{\perp}^{2}}{2 b'^{2}}\right) L_{n}^{|m|}\left(\frac{p_{\perp}^{2}}{b'^{2}}\right) \exp \left(i m \theta\right),
		\end{aligned}
	\end{equation}
	where $b'=\sqrt{x}~b$ is the $x$-dependent HO basis scale parameter~\cite{PhysRevD.111.094012}, $x$ is the longitudinal momentum fraction, and $L_{n}^{|m|}$ is the associated Laguerre polynomial. $n$ and $m$ are the radial quantum number and the angular quantum number, respectively. Here, $p_\perp=|\vec{p}_\perp|$, and $\theta=\arg(\vec{p}_\perp)$. 
	
	The state of the $i$-th parton is fully specified by a set of quantum numbers $\{\alpha_i\}=\{k_{i},n_{i},m_{i},s_{i}\}$, where $s_i$ denotes the light-front helicity. Furthermore, we require that the
	many-body basis states possess a total angular momentum projection,
	\begin{equation}
		M_J=\sum_i (m_i+s_i).
	\end{equation}
	
	As dictated by SU(3) color symmetry, both the $|q\bar{q}\rangle$ and $|q\bar{q}g\rangle$ systems can form only one color-singlet state. Therefore, our basis states are fully specified by their kinematic and spin quantum numbers, without the need for introducing an additional color index.
	
	In order to keep the basis size finite, we introduce two truncation parameters $N_{\max}$ and $K$ in the transverse and longitudinal directions, respectively.
	\begin{equation}
		N_{\max} \ge \sum_{i} (2n_i+|m_i|+1),\ K=\sum_i k_i,
	\end{equation}
	where $K$ specifies the resolution of the longitudinal basis and $N_{\max}$ determines the ultraviolet (UV) and infrared (IR) cutoffs as 
	$\Lambda_{\mathrm{UV}} \simeq b \sqrt{N_{\max }}$ and $\lambda_{\mathrm{IR}} \simeq b / \sqrt{N_{\max }}$~\cite{ZHAO201465}.
	
	The LFWFs in the momentum space are written as a sum over contributions from individual Fock sectors. The contribution from each Fock sector is expressed as,
	\begin{equation}\label{eqn:wave}
		\begin{aligned}
			& \Psi_{\mathcal{N}}^{M_J,\left\{s_i\right\}}\left(\left\{x_i, \vec{p}_{\perp i}\right\}\right) \\
			& =\sum_{\left\{n_i m_i\right\}} \Psi_{\mathcal{N}}^{M_J}\left(\left\{\alpha_i\right\}\right) \prod_{i=1}^{\mathcal{N}} \phi_{n_i m_i}\left(b', \vec{p}_{\perp i}\right),
		\end{aligned}
	\end{equation}
	where $\Psi^{M_J}_{\mathcal{N}=2}(\{\alpha_i\})$ and $\Psi^{M_J}_{\mathcal{N}=3}(\{\alpha_i\})$ are the LFWFs associated with the $|q\bar{q}\rangle$ and $|q\bar{q}g\rangle$ Fock sectors, respectively. These wave functions are obtained in the BLFQ basis by diagonalizing Eq.~\eqref{fc} numerically.

	\subsection{Regularization and Renormalization}
	
	In the heavy meson systems, the quark and antiquark interact through the exchange of dynamical gluons and the instantaneous gluon interaction. In the full theory, rotational symmetry ensures a delicate cancellation of ultraviolet (UV) behaviors between the interactions exchanging the dynamical gluon and instantaneous gluon~\cite{BRODSKY1998299}. However, this cancellation is broken by the basis truncation in BLFQ. Specifically, while the transverse momentum of the dynamical gluon is limited by the basis truncation, there are no corresponding truncations applied to the instantaneous gluon. This mismatch leads to artificial divergences and breaks the rotational symmetry~\cite{Tang:1991rc}.
	
	To regulate the mismatch in the transverse momentum between the dynamical and the instantaneous gluon, we apply a multiplicative regulating function $e^{-p_\perp^2/b_{\text{inst}}^2}$ to the instantaneous gluon exchange interaction \cite{zhao:2014,Lan:2021wok}, where $p_\perp$ is the transverse momentum of the ``instantaneous gluon". The parameter $b_{\text{inst}}$ characterizes the regularization scale in the transverse momentum of the ``instantaneous gluon".
	This regulating function balances the UV cutoffs between the instantaneous and dynamical gluon interactions and thus helps maintain the rotational symmetry of the heavy meson system. 
	
	In the $|q\bar{q}\rangle$ Fock sector, the quark and antiquark masses are renormalized by the dynamical gluon. Following the Fock-sector dependent renormalization scheme~\cite{Tang:1991rc,PhysRevD.77.085028,PhysRevD.86.085006}, the mass counterterm ($\delta m_q=m_{q0}-m_q$) is introduced to compensate for the dynamical-gluon-induced self-energy correction~\cite{ZHAO201465}. It is determined by matching the eigenvalue of the single-quark system to the renormalized quark mass $m_q$~\cite{zhao:2014}.

	In order to model the nonperturbative physics from the truncated higher Fock sectors, we introduce two effective mass parameters~\cite{Lan:2021wok}: First, following the analysis in Ref.~\cite{PhysRevD.58.096015,3z39-l1kg}, we employ an independent vertex mass $m_{f(\bar{f})}$, distinct from the kinetic mass $m_q$.
	This vertex mass parameterizes the nonperturbative effects in connection with the chiral symmetry breaking in light-front dynamics~\cite{PhysRevD.58.096015}.
	Second, we introduce an effective gluon mass $m_g$ in the kinetic energy term in the $|q\bar{q}g\rangle$ sector. This effective gluon mass simulates the confining effects in the $|q\bar{q}g\rangle$ sector and regulates the infrared divergence in gluon emission~\cite{PhysRevD.108.094002}.
	
	In the future, as the higher Fock sectors will be progressively included in the BLFQ basis, we expect that $m_f$ gradually converges to the kinetic quark mass $m_q$ and the effective gluon mass $m_g$ approaches zero as required by the gauge invariance.
	
	\section{NUMERICAL RESULTS}\label{sec3}
	Our calculation involves two types of parameters: the truncation parameters $\{N_{\max}, K\}$ and the input parameters in the Hamiltonian $\{\alpha_s, m_q, m_f, m_g, b, b_{\rm inst}, \kappa_T, \kappa_L\}$.
	We perform the calculations with fixed basis truncations of $N_{\max}=12$ and $K=17$. The strong coupling constant is estimated from the mass scale of the heavy quarkonium system~\cite{y.li4}, yielding $\alpha_s^{b\bar{b}}=0.23$ and $\alpha_s^{c\bar{c}}=0.30$. Considering the rotational symmetry in the nonrelativistic limit, we set $\kappa_T=\kappa_L=b$ for expedited convergence~\cite{y.li4,y.li1}. The remaining parameters $\{m_q,m_f,m_g,b,b_{\rm inst}\}$ are obtained by fitting the mass spectra of the low-lying charmonium and bottomonium states. The resulting parameter values are summarized in Table~\ref{parameter}.
	For $\rm B_c$ mesons, the strong coupling constant $g$, the scale parameter $b_\mathrm{inst}$ in the instantaneous interaction, and the basis scale $b$ are taken from the root-mean-square (rms) of the corresponding parameters for the charmonium and bottomonium systems~\cite{PhysRevD.95.034016,s.tang},
	\begin{equation}
		\begin{aligned}
			g^{b \bar{c}}=&\sqrt{\frac{(g^{c \bar{c}})^2+(g^{b \bar{b}})^2}{2}}, \\ 
			b^{b \bar{c}}=&\sqrt{\frac{(b^{c \bar{c}})^2+(b^{b \bar{b}})^2}{2}}, \\
			b^{b\bar{c}}_{\mathrm{inst}}=&\sqrt{\frac{(b^{c\bar{c}}_{\mathrm{inst}})^2+(b^{b\bar{b}}_{\mathrm{inst}})^2}{2}}.
		\end{aligned}
	\end{equation}
	The superscripts $c\bar{c}$, $b\bar{b}$, and $b\bar{c}$ denote the charmonium,  bottomonium, and $\rm B_c$ meson systems, respectively. Note that there are no new parameters for the $\rm B_c$ meson system.
	
	\begin{table*}
		\centering
		\caption{The input parameters of the heavy meson systems with truncation parameters: $N_{\max}=12,K=17$.}
		\begin{tabular}{c*{9}{>{\centering\arraybackslash}p{5em}}}
			\toprule
			\ &$m^c_{q}{\rm (GeV)}$&$m^b_{q}{\rm (GeV)}$&$\alpha_s$&\parbox{5.7em}{$b=\kappa_T=\kappa_L$\\${\rm (GeV)}$}&$m^c_{f}{\rm (GeV)}$&$m^b_{f}{\rm (GeV)}$&$m_g{\rm (GeV)}$&$b_\mathrm{inst}{\rm (GeV)}$\\
			\midrule
			charmonium&		1.54&		$\cdots$&	0.30&	1.23&	5.04&		$\cdots$&	0.50&	3.99\\
			bottomonium&	$\cdots$&	4.78&		0.23&	1.84&	$\cdots$&	13.20&		0.50&	7.35\\
			$\rm B_c$ meson&	1.54&		4.78&		0.26&	1.56&	5.04&		13.20&		0.50&	5.91\\
			\bottomrule
			\label{parameter}
		\end{tabular}
	\end{table*}
	\begin{figure*}
		\centering
		\subfloat[charmonium states]{\includegraphics[width=0.31\linewidth]{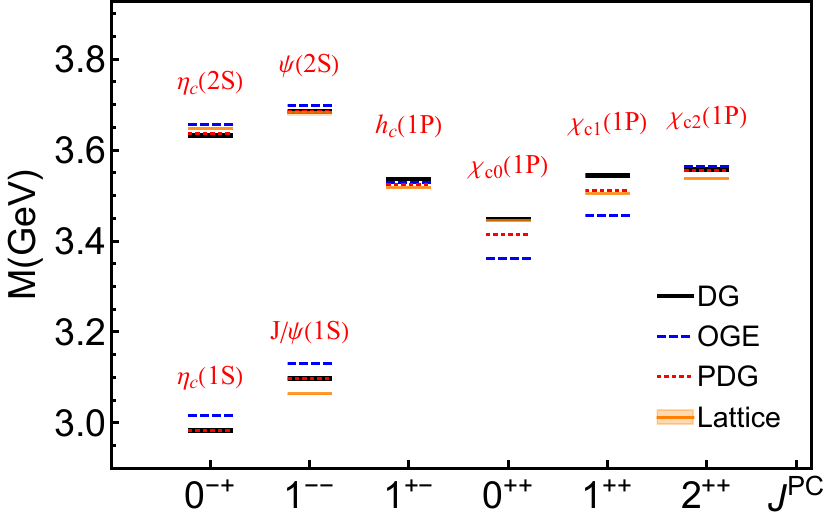}}\quad
		\subfloat[bottomonium states]{\includegraphics[width=0.321\linewidth]{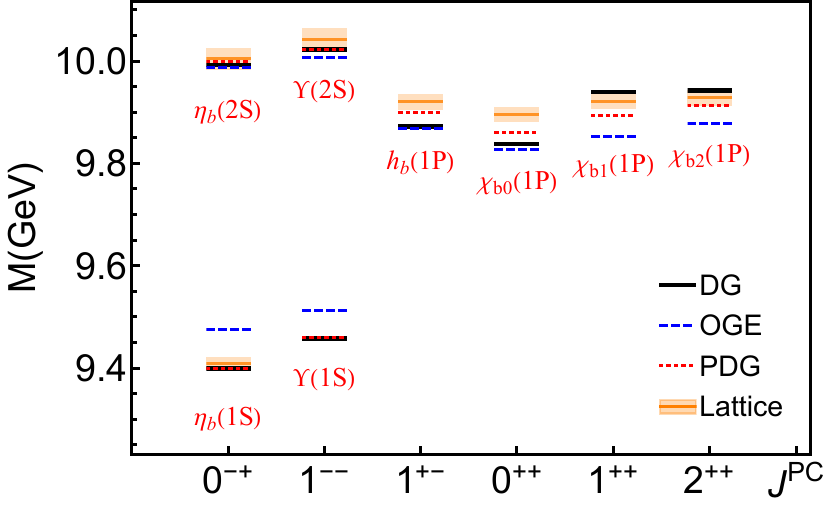}}\quad
		\subfloat[$\rm B_c$ meson states]{\includegraphics[width=0.31\linewidth]{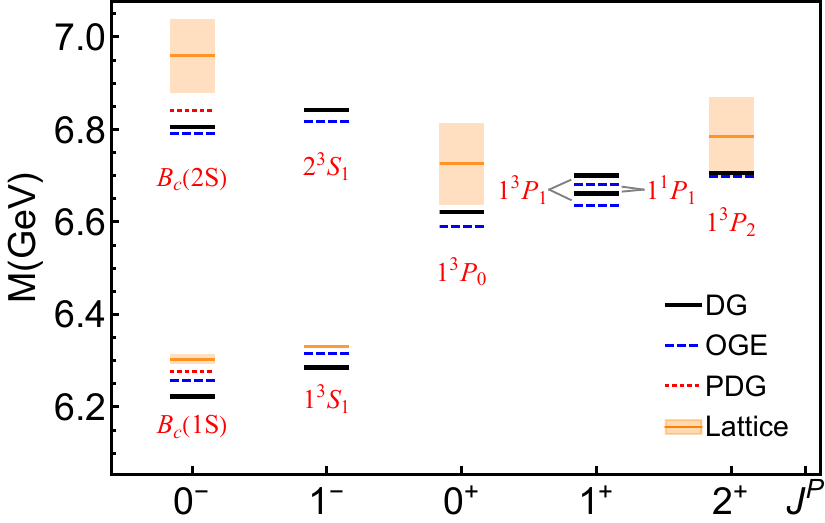}}
		\caption{The reconstructed mass spectrum of charmonium, bottomonium and $\rm B_c$ meson states from the $M_J=0$ sector. The horizontal axis labels the quantum numbers $J^{PC}$, and the vertical axis is invariant mass in GeV. We denote the BLFQ calculation including the $|q\bar{q}\rangle+|q\bar{q}g\rangle$ sectors as Dynamical Gluon (DG), to distinguish it from the BLFQ results based on the One-Gluon Exchange (OGE) effective interactions. The black lines are the results from BLFQ DG, and we compare with data from particle data group (PDG)~\cite{pdg} (dotted red line), BLFQ OGE with truncation parameters $N_{\max} = L_{\max} = 32$~\cite{y.li1,s.tang} (dashed blue line) and the results from Lattice QCD~\cite{HadronSpectrum:2012gic,Aarts:2014cda,PhysRevLett.94.172001,PhysRevLett.104.022001,DAVIES1996131} (orange band). The PDG values serve as the inputs to fix the model parameters. For charmonium and bottomonium systems, the results are marked by boxes representing the spreads of the mass eigenvalues in different $M_J$ sectors. For $\rm B_c$ meson, only the results in $M_J =0$ sector are shown due to the difficulty in identifying the multiplets with the same $J$, see text for details. }
		\label{mass}
	\end{figure*}

	\subsection{Mass Spectrum}
	In this work, we calculate eight states below the respective dissociation thresholds for the charmonium, bottomonium, and $\rm B_c$ meson states.
	We determine the total angular momentum quantum number $J$ of the eigenstates by identifying the multiplets with different $M_J$ belonging to the same $J$. Because on the light front the rotational symmetry is dynamical and can be broken by Fock space truncations, the degeneracy between the eigenstates in the same multiplet is not exact. Therefore, in order to identify the eigenstates, we additionally take into account the parity $P$ and the charge conjugation parity $C$ for the heavy meson states.
	Since the true parity symmetry is also dynamical on the light front, we utilize a kinematic substitute known as the mirror parity $\hat{m}_{P}=\hat{R}_{x}(\pi) P$~\cite{PhysRevD.73.036007}. 
	In this work, the mirror parity $m_P$ and charge conjugation $C$ are evaluated in terms of the light-front wave functions as~\cite{y.li4,PhysRevD.45.3755,Trittmann:1997xz,PhysRevD.73.036007},
	
	\begin{equation}
		\begin{aligned}
			m_P&=(-i)^{2J}P=\langle\Psi^{-M_J}|\hat{m}_{p}|\Psi^{M_J}\rangle\\
			&=\sum_{\mathcal{N},\{\alpha_i\}}(-1)^{m} \Psi_\mathcal{N}^{-M_J *}(\{k_i,n_i,-m_i,-s_i\})\\ &\qquad\qquad\qquad*\Psi_\mathcal{N}^{M_J}(\{k_i,n_i,m_i,s_i\}), \\
			\\
			C&=\langle\Psi^{M_J}|\hat{C}|\Psi^{M_J}\rangle\\
			&=\sum_{\mathcal{N},\{\alpha_i\}}(-1)^{\mathcal{N}-1} \Psi_\mathcal{N}^{M_J *}(\{\alpha_i'\}) \Psi_\mathcal{N}^{M_J}(\{\alpha_i\}),
		\end{aligned}
	\end{equation}
	where $\hat{P}$ and $\hat{C}$ denote the parity and charge conjugation operators, respectively. Here $m=\sum_i m_i$ denotes the projection of total angular momentum, and $J$ denotes the total angular momentum of the heavy meson states. Once both the mirror parity $m_p$ and the total angular momentum $J$ are known, the true parity can be obtained from $m_P=(-i)^{2J}P$. For the charge conjugation $\hat{C}$, the notation $\{\alpha_i'\}$ implies an exchange of the quantum numbers associated with the quark ($i=1$) and antiquark ($i=2$) in the LFWFs, since the charge conjugation operator transforms particles into antiparticles and vice-versa. Furthermore, the phase factor $(-1)^{\mathcal{N}-1}$ is introduced to account for the intrinsic $C$-parity of $-1$ carried by the gluon in the $|q\bar{q}g\rangle$ Fock sector.
	
	By fitting the obtained eigenvalues to the experimental results from the Particle Data Group (PDG)~\cite{pdg}, we determine the parameters in our input Hamiltonian as listed in Table~\ref{parameter}.
	The obtained spectra for the eight lowest-lying charmonium, bottomonium, and $\rm B_c$ meson states are compared to the BLFQ OGE and lattice QCD results as well as the experimental data from the PDG in Fig.~\ref{mass}. Our results with dynamical gluon (BLFQ DG) are generally consistent with the experimental data from PDG. Compared with BLFQ OGE~\cite{y.li1,s.tang}, our results show improvements on the masses of both the P-wave and S-wave states.
	We calculate the rms deviation between the masses of our lowest 8 states and the PDG data. For the charmonium (bottomonium) system, the obtained rms mass deviation is 17.24 MeV (23.40 MeV) in the $M_J=0$ sector. As a comparison, the earlier BLFQ OGE calculation yields 31 MeV (38 MeV) for the charmonium (bottomonium) system~\cite{y.li1}, where the lowest 8 (14) states below the $D\bar{D}$ ($B\bar{B}$) threshold are considered.

	The dashed lines indicate the rms of invariant masses, which are estimated from the obtained eigenvalues for different $M_J$ components,
	\begin{equation}
		\overline{M} \equiv \sqrt{\frac{\sum_{M_J=-J}^{J}M^2(M_J)}{2J+1}},
	\end{equation}
	where $M(M_J)$ is the mass eigenvalue of the state with total angular momentum projection $M_J$. The boxes show the mass spreads $\delta M_{J} \equiv \max \left[M(M_{J})\right]-\min \left[M(M_{J})\right]$. Since $\rm B_c$ mesons do not have well-defined charge conjugation parity, state identification is more challenging, especially for the $M_J \neq 0$ components. Therefore, for $\rm B_c$ mesons, we only present the mass eigenvalues from the $M_J=0$ component in Fig.~\ref{mass}.
	
	\subsection{Light-front Wave Functions}
	In our truncated basis, the eigenstates of heavy mesons are considered as a superposition of the $|q\bar{q}\rangle$ and $|q\bar{q}g\rangle$ sectors. We define $N_{q\bar{q}}$ as the norm of the leading sector, such that $N_{q\bar{q}}^2$ yields the probability from the leading Fock sector, 
	\begin{equation}
		N_{q\bar{q}}^2 = \sum_{\{s_{i}\}} \int \left[\mathrm{d}^3p\right]_2  \big|\Psi^{\{s_{i}\}}_2\left(\{x_i,\vec{p}_{\perp i}\}\right)\big|^2,
		\label{noc}
	\end{equation}
	with
	\begin{equation}
		[\mathrm{d}^3p]_n \equiv \prod_{i=1}^n \left[\frac{\mathrm{d}x_i \mathrm{d}^2\vec{p}_{\perp i}}{16\pi^3}\right] 16\pi^3 \delta\left(1-\sum_{i=1}^n x_i\right) \delta^2\left(\sum_{i=1}^n \vec{p}_{\perp i}\right).
	\end{equation}

	\begin{table}[H]
		\centering
		\caption{The mass spectrum, the leading Fock sector probability, $N_{q\bar{q}}^2$, and the contributions from different spin components for heavy mesons in the $M_J=0$ sector.}
		\begin{tabular}{c*{5}{>{\centering\arraybackslash}p{4.2em}}}
			\toprule
			\ &$M{\rm (GeV)}$&$N_{q\bar{q}}^2$&$\downarrow\downarrow(\uparrow\uparrow)$&$\downarrow\uparrow-\uparrow\downarrow$&$\downarrow\uparrow+\uparrow\downarrow$\\
			\midrule
			$\eta_c(\mathrm{1S})$	&2.983&0.643&0.004&0.636&0.000\\
			$J/\psi(\mathrm{1S})$	&3.096&0.639&0.000&0.000&0.639\\
			$\chi_{c0}(\mathrm{1P})$&3.448&0.533&0.042&0.000&0.447\\
			$\chi_{c1}(\mathrm{1P})$&3.543&0.501&0.244&0.014&0.000\\
			$h_c(\mathrm{1P})$		&3.537&0.531&0.000&0.530&0.000\\
			$\chi_{c2}(\mathrm{1P})$&3.557&0.503&0.208&0.000&0.087\\
			$\eta_c(\mathrm{2S})$	&3.632&0.504&0.002&0.499&0.000\\
			$\psi(\mathrm{2S})$		&3.685&0.529&0.001&0.000&0.527\\
			\midrule
			$\eta_b(\mathrm{1S})$	&9.399&0.811&0.002&0.808&0.000\\
			$\Upsilon(\mathrm{1S})$	&9.458&0.815&0.000&0.000&0.815\\
			$\chi_{b0}(\mathrm{1P})$&9.838&0.713&0.011&0.000&0.691\\
			$\chi_{b1}(\mathrm{1P})$&9.940&0.690&0.338&0.014&0.000\\
			$h_b(\mathrm{1P})$		&9.872&0.711&0.000&0.711&0.000\\
			$\chi_{b2}(\mathrm{1P})$&9.942&0.690&0.333&0.000&0.023\\
			$\eta_b(\mathrm{2S})$	&9.992&0.688&0.001&0.686&0.000\\
			$\Upsilon(\mathrm{2S})$&10.023&0.698&0.000&0.000&0.698\\
			\midrule
			$\rm B_c(\mathrm{1S})$	&6.222&0.724&0.003&0.717&0.000\\
			$1^3\mathrm{S_1}$		&6.285&0.722&0.001&0.000&0.721\\
			$1^3\mathrm{P_0}$		&6.621&0.629&0.029&0.000&0.571\\
			$1^3\mathrm{P_1}$		&6.662&0.629&0.286&0.034&0.000\\
			$1^1\mathrm{P_1}$		&6.701&0.605&0.010&0.609&0.000\\
			$1^3\mathrm{P_2}$		&6.706&0.605&0.272&0.000&0.062\\
			$\rm B_c(\mathrm{2S})$	&6.806&0.591&0.002&0.587&0.000\\
			$2^3\mathrm{S_1}$		&6.842&0.604&0.001&0.000&0.602\\
			\bottomrule
			\label{norm}
		\end{tabular}
	\end{table}
	
	We list the norm of the $|q\bar{q}\rangle$ sector, $N^2_{q\bar{q}}$  for the eight lowest-lying states in the $M_J=0$ sector in Table~\ref{norm}. As shown in Table~\ref{norm}, the leading Fock sector is dominant for all low-lying heavy meson states, with its probabilities $N_{q\bar{q}}^2$ varying between 0.5 and 1. We observe that within each of the three systems $N_{q\bar{q}}^2$ is positively correlated with the mass of the heavy quarks, that is, the bottomonium states have a larger probability to stay in the $|q\bar{q}\rangle$ Fock sector compared to the charmonium and $B_c$ states. This behavior is expected as gluon radiation is suppressed by the mass of the heavy quarks. Within the same heavy meson system, $N^2_{q\bar{q}}$ decreases as the heavy meson mass increases, suggesting that compared to the ground states, the excited states have a larger probability to stay in the $|q\bar{q}g\rangle$ Fock sector.
	
\begin{figure*}
		\centering
		\setcounter{subfigure}{16}
		\subfloat[$\rm B_c(\mathrm{1S})$]{\includegraphics[width=0.45\linewidth]{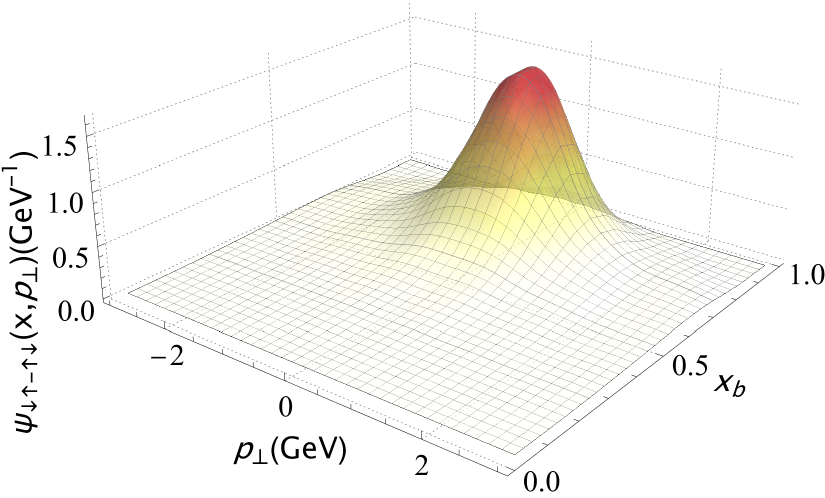}}\quad
		\subfloat[$1^3\mathrm{S_1}$]{\includegraphics[width=0.45\linewidth]{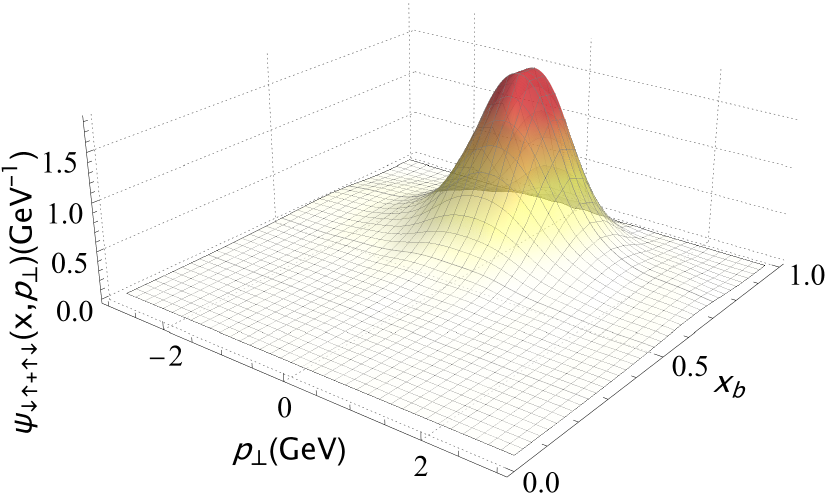}}\quad
		\subfloat[$1^3\mathrm{P_0}$]{\includegraphics[width=0.45\linewidth]{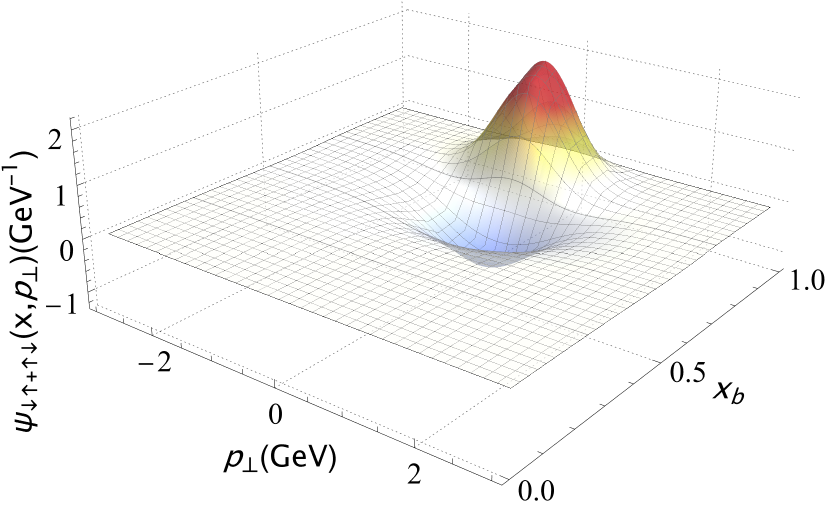}}\quad
		\subfloat[$1^3\mathrm{P_1}$]{\includegraphics[width=0.45\linewidth]{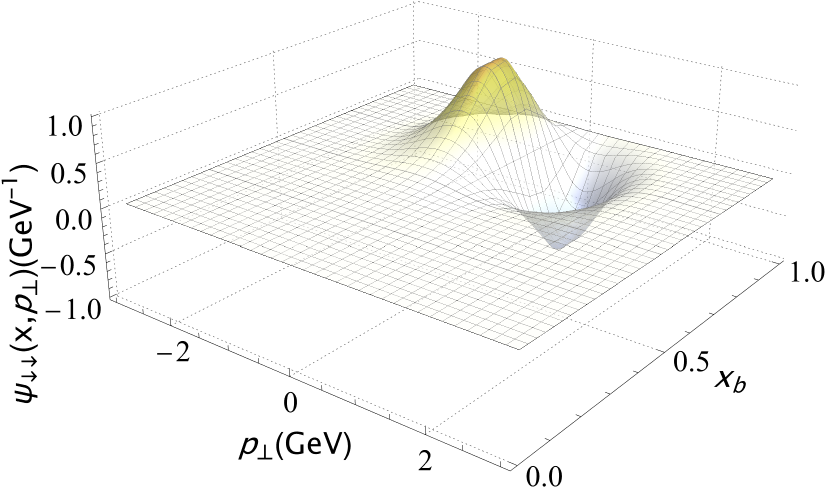}}\quad
		\subfloat[$1^1\mathrm{P_1}$]{\includegraphics[width=0.45\linewidth]{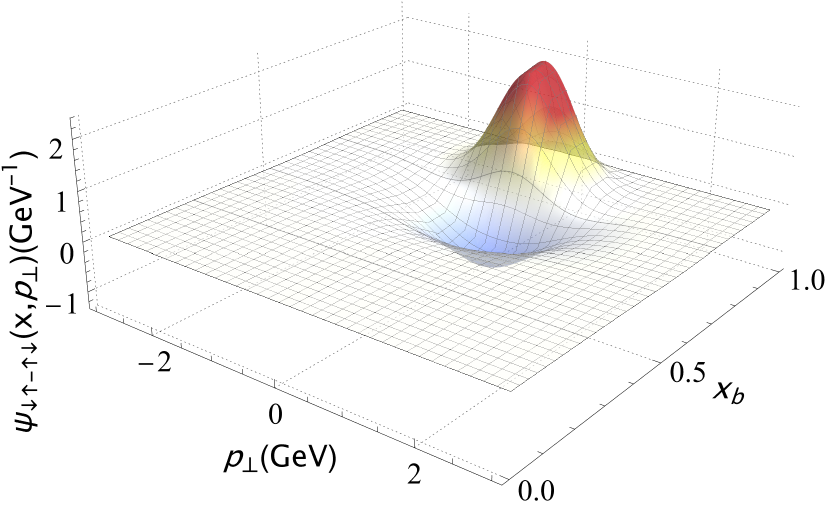}}\quad
		\subfloat[$1^3\mathrm{P_2}$]{\includegraphics[width=0.45\linewidth]{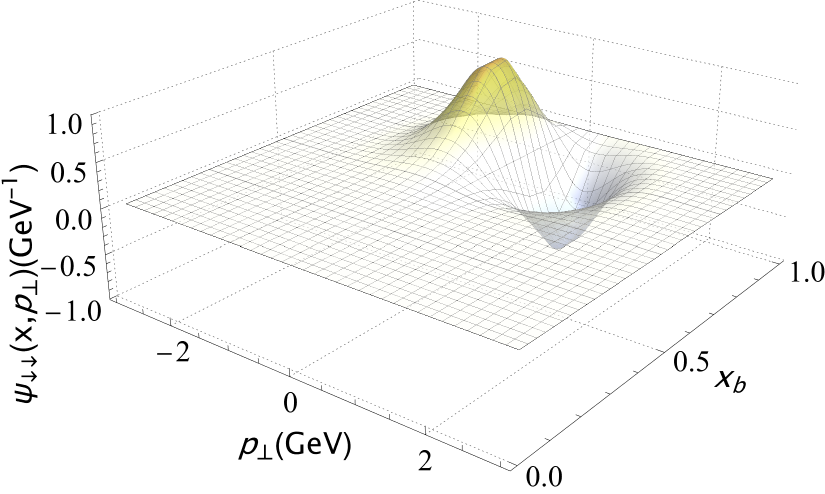}}\quad
		\subfloat[$\rm B_c(\mathrm{2S})$]{\includegraphics[width=0.45\linewidth]{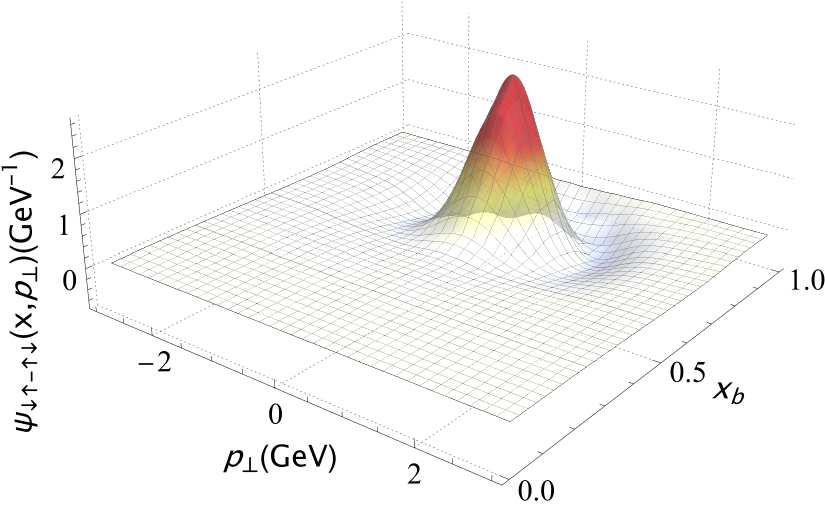}}\quad
		\subfloat[$2^3\mathrm{S_1}$]{\includegraphics[width=0.45\linewidth]{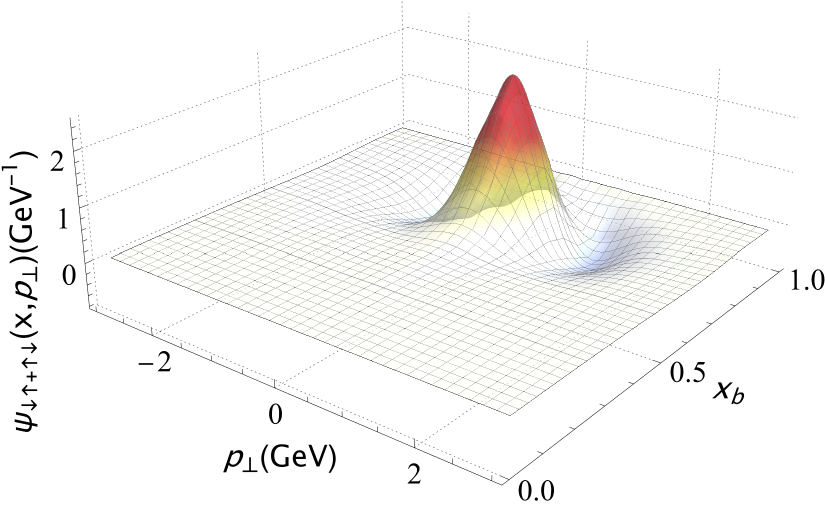}}
		\caption{Normalized momentum-space LFWFs in the $|q\bar{q}\rangle$ Fock sector for heavy meson states with $M_J=0$. Panels (a)-(h): the charmonium states; panels (i)-(p): the bottomonium states; and panels (q)-(x): the $\rm B_c$ meson states. For $\rm B_c$ meson states, $x_b$ is the longitudinal momentum fraction of the $b$ quark. See text for details on normalization.}
		\label{wave}
	\end{figure*}

	To visualize the LFWFs in the leading Fock sector, we first normalize them by their norm $N_{q\bar{q}}$ and then transform the LFWFs from the single particle coordinate into the relative coordinate by the Talmi-Moshinsky (TM) transformation\cite{y.li1,ChaosCador2004CommonGF,PhysRevD.75.094016}. Next we consider the intrinsic part $\Psi^{s_1 s_2}_2(x,\vec{p}_\perp)$, where $\vec{p}_\perp=x_2\vec{p}_{\perp 1}-x_1\vec{p}_{\perp 2}$ is the relative transverse momentum between $q$ and $\bar{q}$.
	The angular dependence of the LFWFs in the transverse direction factorizes as $\Psi^{s_1s_2}_2(x, \vec{p}_\perp)=\psi^{s_1s_2}_2(x, p_\perp)e^{\mathrm{i} m \theta}$, with $p_\perp=|\vec{p}_\perp|$, $\theta=\arg(\vec{p}_\perp)$, and $m= m_1+m_2$ being the orbital angular momentum projection on the transverse plane.
	There are four possible spin combinations ($\downarrow\downarrow,\uparrow\uparrow,\downarrow\uparrow-\uparrow\downarrow,\downarrow\uparrow+\uparrow\downarrow$) in each mass eigenstate, where $\uparrow(\downarrow)$ denotes the quark (antiquark) helicity and $\psi_{\downarrow\uparrow\pm\uparrow\downarrow}=(\psi_{\downarrow\uparrow}\pm\psi_{\uparrow\downarrow})/\sqrt{2}$. The probabilities of these components for the $M_J=0$ states are listed in Table~\ref{norm}. In each mass eigenstate, there is one spin combination dominant over other possible combinations. Due to the mirror symmetry of the meson states, the probabilities for the $\downarrow\downarrow$ and $\uparrow\uparrow$ configurations are identical for $M_J=0$ components, so they are listed together in the column $\downarrow\downarrow(\uparrow\uparrow)$.
	Following Ref.~\cite{y.li1}, we plot the LFWFs of the dominant spin component of the $M_J=0$ states as a function of $x$ and $p_\perp$ in Fig.~\ref{wave} by dropping the phase $e^{\mathrm{i} m \theta}$ and retaining the relative sign between positive and negative $p_\perp$,
	
	\begin{equation}
		\psi^{s_1s_2}_2\left(x, p_{\perp}\right) \to \left\{
		\begin{array}{ll}
			\frac{1}{N_{q \bar{q}}} \psi^{s_1s_2}_2\left(x, p_{\perp}\right), & p_{\perp} \geq 0, \\
			\frac{1}{N_{q \bar{q}}} \psi^{s_1s_2}_2\left(x, -p_{\perp}\right) \times(-1)^{m}, & p_{\perp}<0.
		\end{array}\right.
	\end{equation}
	
	For S-wave charmonium and bottomonium states, the leading $|q\bar{q}\rangle$ Fock sector does not carry orbital angular momentum, resulting in an azimuthally symmetric LFWF in the transverse plane. Compared to the ground (1S) states, the excited (2S) states exhibit a nodal structure in both transverse and longitudinal directions. We note that the nodal structure in the longitudinal direction is less pronounced compared to the transverse directions, which indicates the absence of explicit rotational symmetry between the transverse and longitudinal directions on the light front.
	
	Compared to the charmonium states, the bottomonium states exhibit a broader distribution in the transverse momentum space due to their larger mass scale, which also leads to smaller spatial sizes for the bottomonium states, as discussed in Sec.~\ref{eff}.
	
	\begin{figure*}
		\centering
		\setcounter{subfigure}{0}
		\subfloat[$\eta_c(\mathrm{1S})$]{\includegraphics[width=0.42\linewidth]{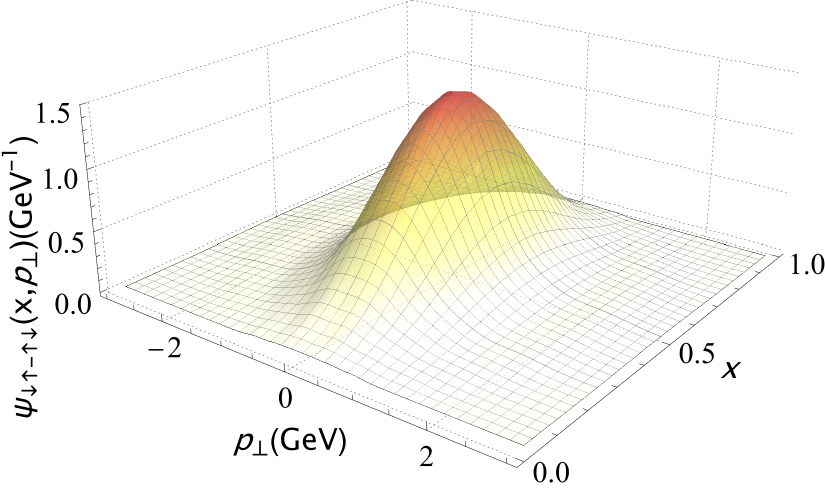}}
		\subfloat{\includegraphics[width=0.07\linewidth]{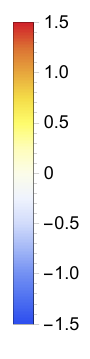}}\ \ 
		\subfloat[$J/\psi(\mathrm{1S})$]{\includegraphics[width=0.42\linewidth]{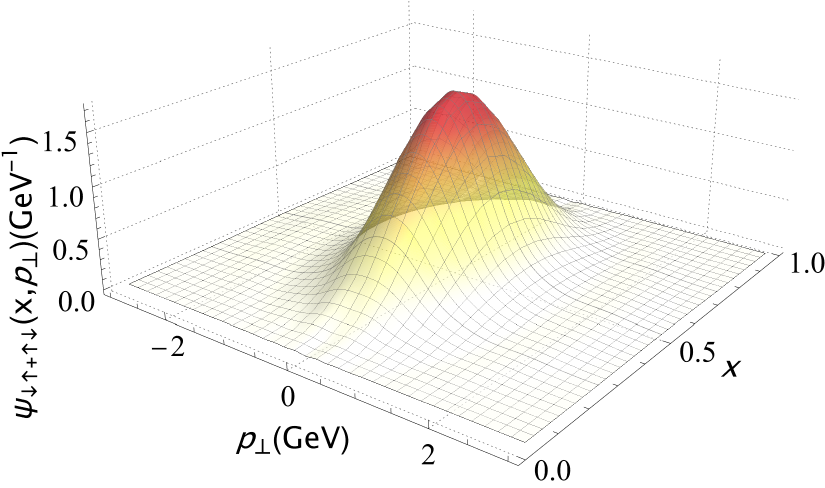}}\quad
		\subfloat[$\chi_{c0}(\mathrm{1P})$]{\includegraphics[width=0.45\linewidth]{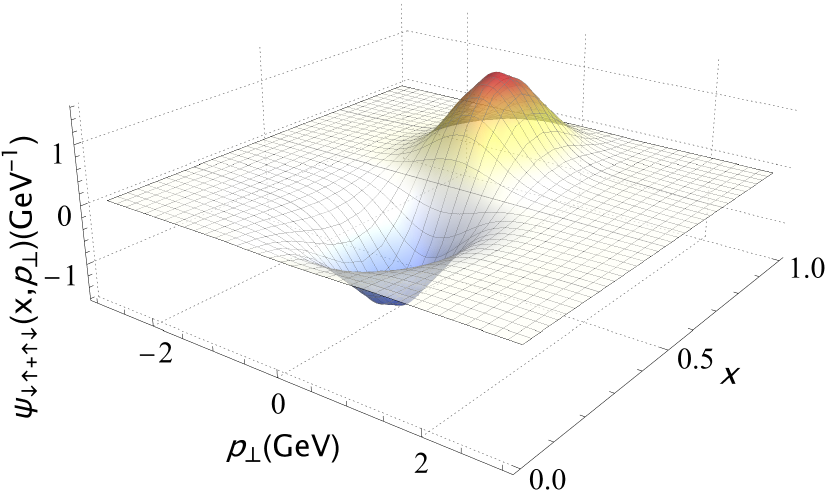}}\quad
		\subfloat[$\chi_{c1}(\mathrm{1P})$]{\includegraphics[width=0.45\linewidth]{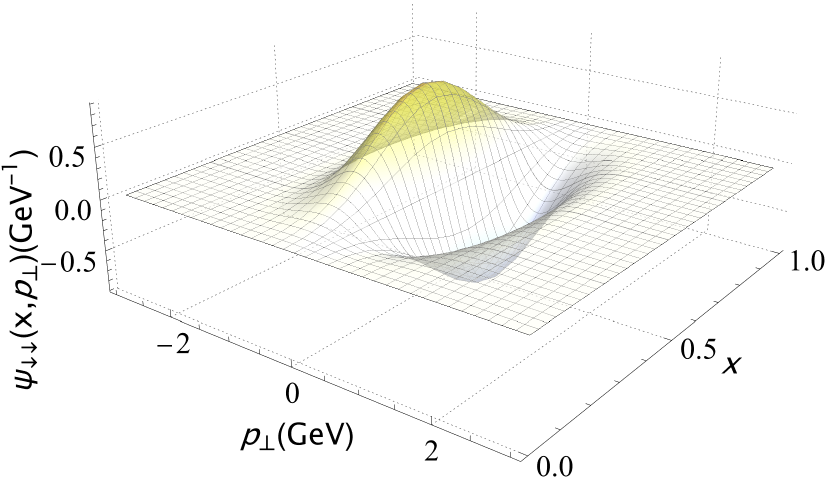}}\quad
		\subfloat[$h_c(\mathrm{1P})$]{\includegraphics[width=0.45\linewidth]{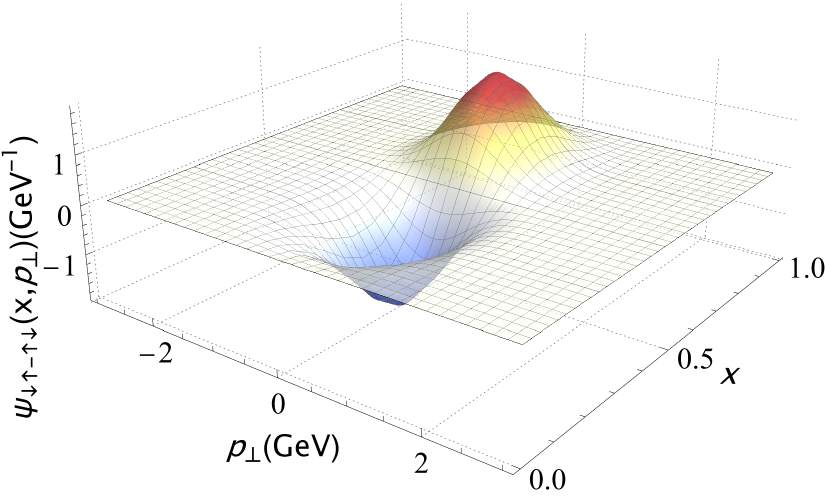}}\quad
		\subfloat[$\chi_{c2}(\mathrm{1P})$]{\includegraphics[width=0.45\linewidth]{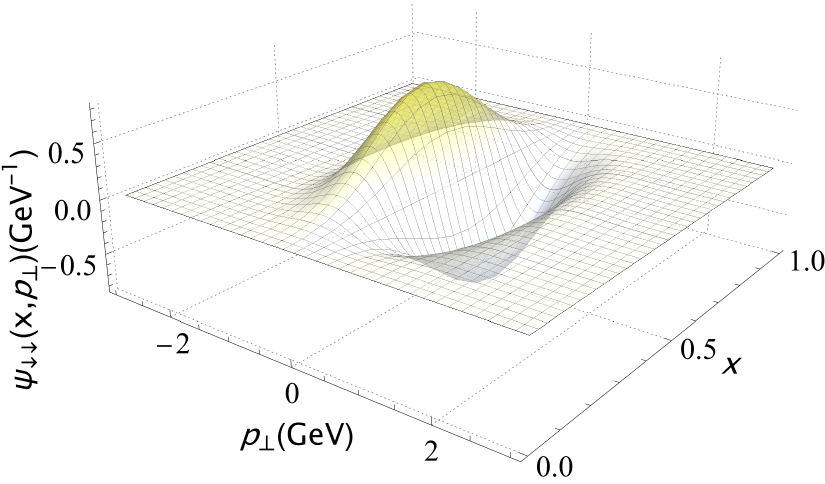}}\quad
		\subfloat[$\eta_c(\mathrm{2S})$]{\includegraphics[width=0.45\linewidth]{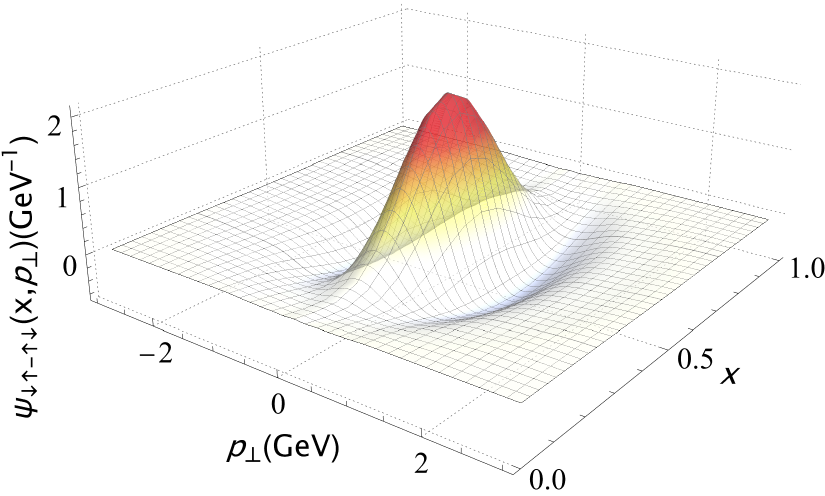}}\quad
		\subfloat[$\psi(\mathrm{2S})$]{\includegraphics[width=0.45\linewidth]{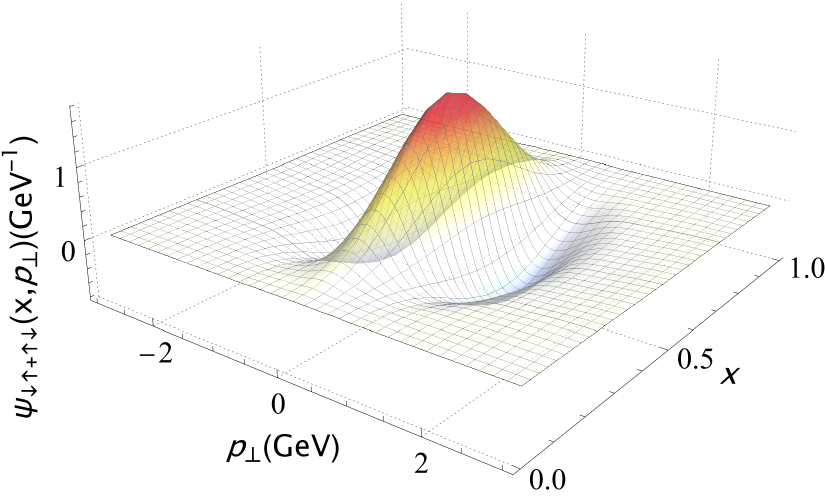}}
	\end{figure*}
	\begin{figure*}                              
		\centering
		\setcounter{subfigure}{8}
		\subfloat[$\eta_b(\mathrm{1S})$]{\includegraphics[width=0.45\linewidth]{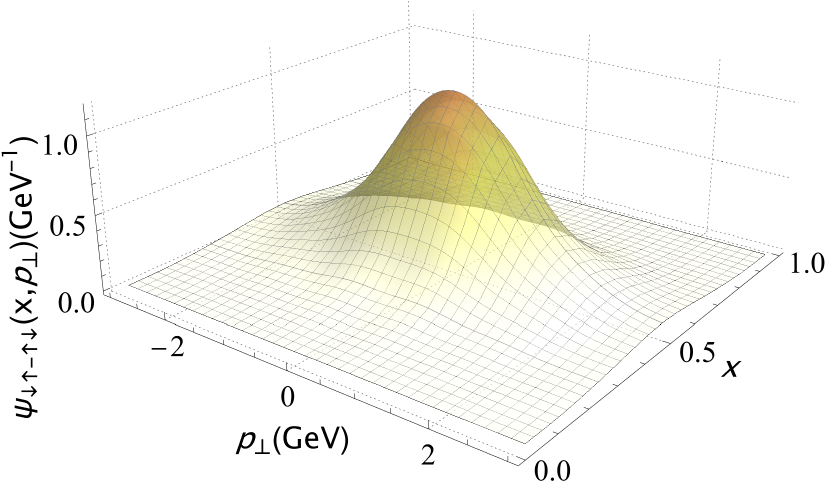}}\quad
		\subfloat[$\Upsilon(\mathrm{1S})$]{\includegraphics[width=0.45\linewidth]{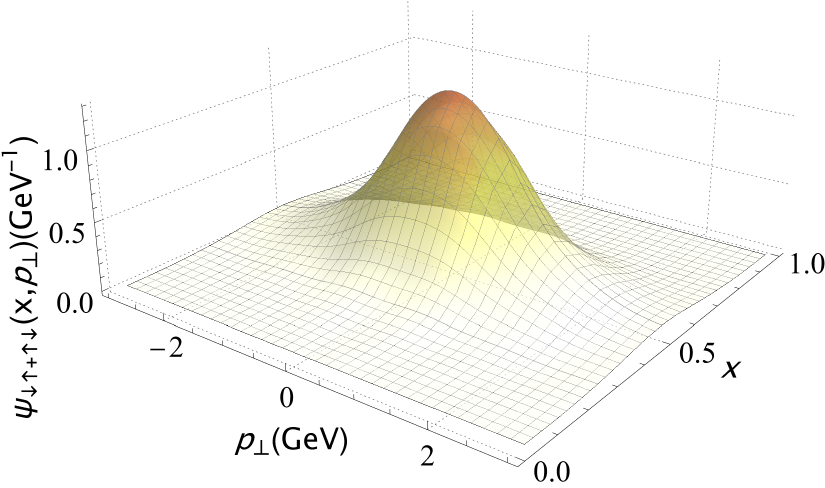}}\quad
		\subfloat[$\chi_{b0}(\mathrm{1P})$]{\includegraphics[width=0.45\linewidth]{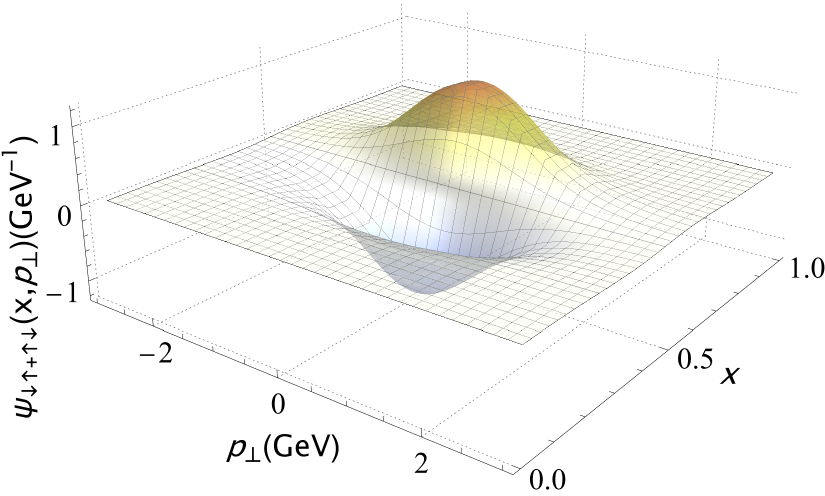}}\quad
		\subfloat[$\chi_{b1}(\mathrm{1P})$]{\includegraphics[width=0.45\linewidth]{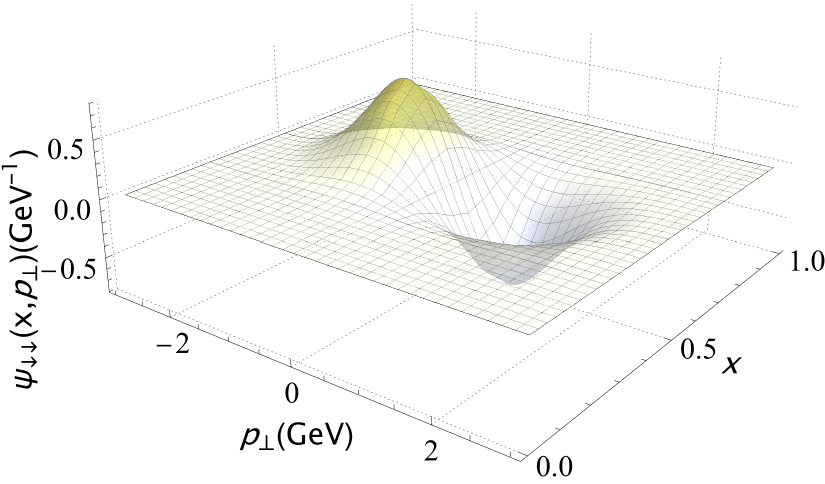}}\quad
		\subfloat[$h_b(\mathrm{1P})$]{\includegraphics[width=0.45\linewidth]{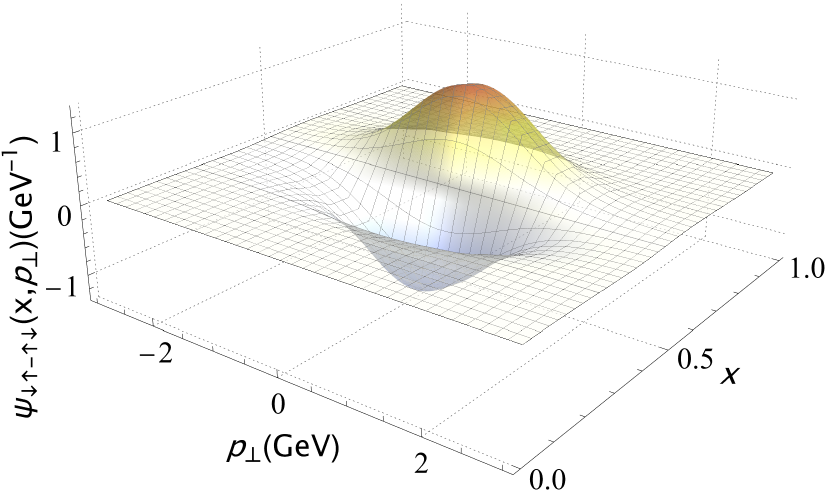}}\quad
		\subfloat[$\chi_{b2}(\mathrm{1P})$]{\includegraphics[width=0.45\linewidth]{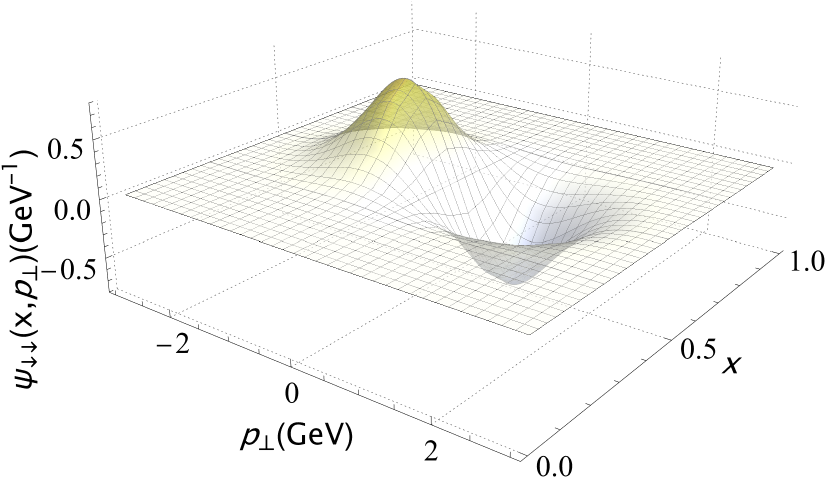}}\quad
		\subfloat[$\eta_b(\mathrm{2S})$]{\includegraphics[width=0.45\linewidth]{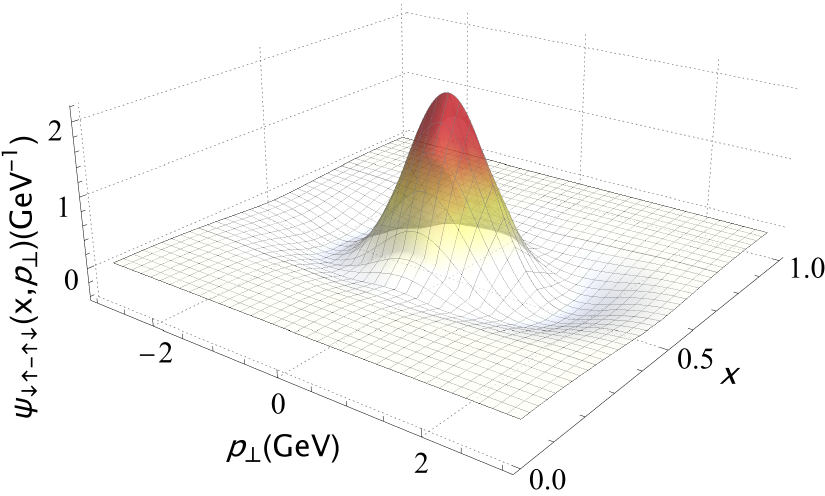}}\quad
		\subfloat[$\Upsilon(\mathrm{2S})$]{\includegraphics[width=0.45\linewidth]{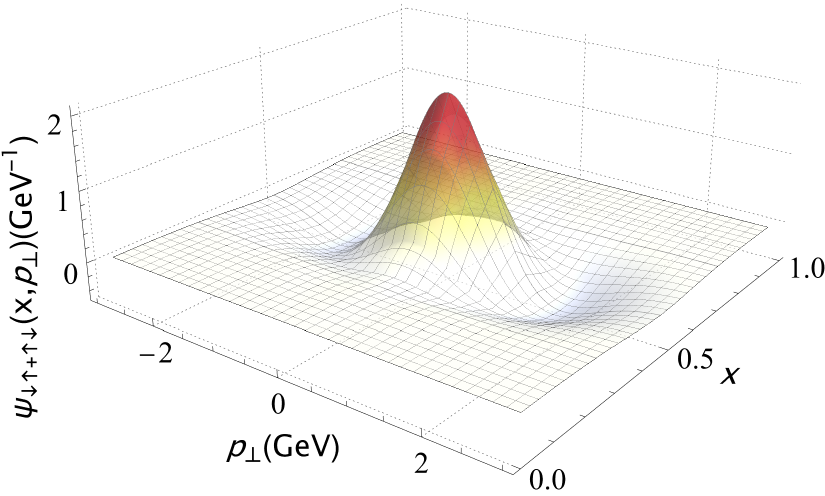}}\\
	\end{figure*}
	
	\begin{figure*}
		\centering
		\setcounter{subfigure}{0}
		\subfloat[charmonium states]{\includegraphics[width=0.32\linewidth]{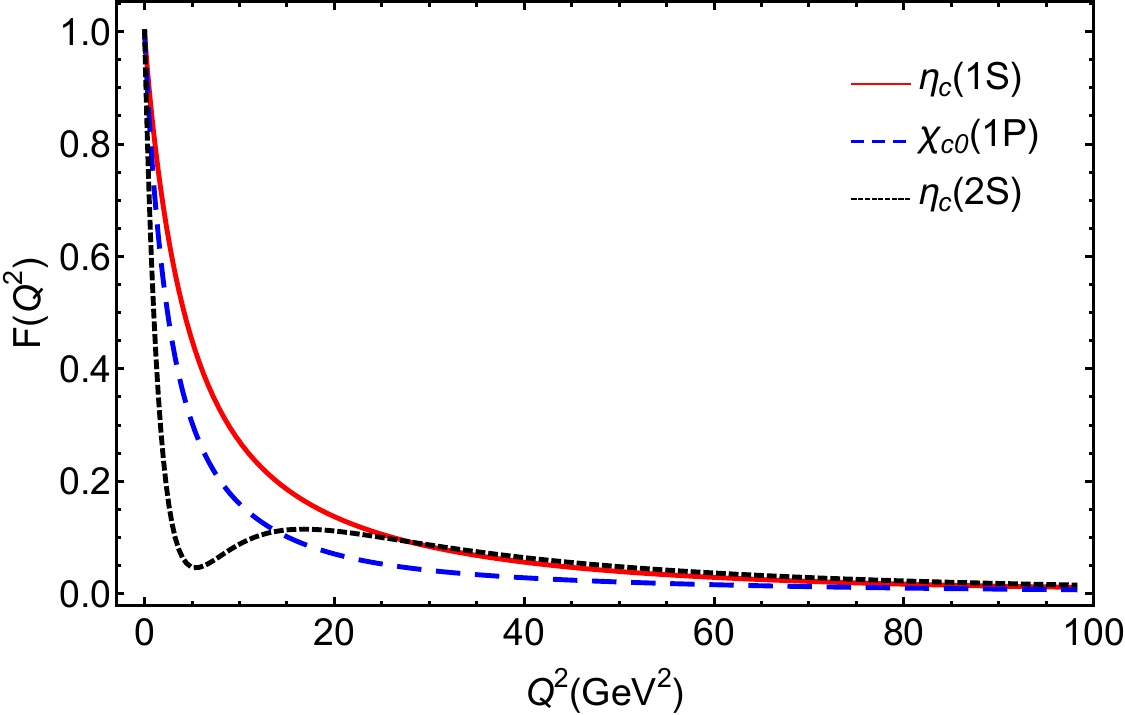}}\quad
		\subfloat[bottomonium states]{\includegraphics[width=0.32\linewidth]{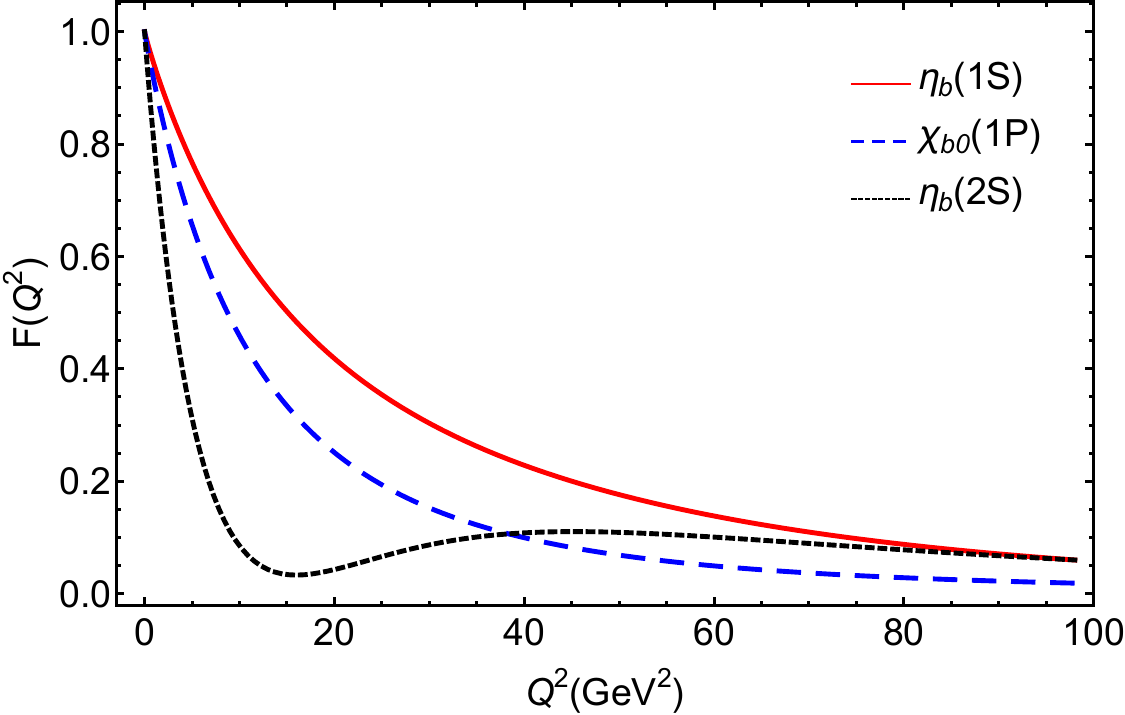}}\quad
		\subfloat[$\rm B_c$ meson states]{\includegraphics[width=0.32\linewidth]{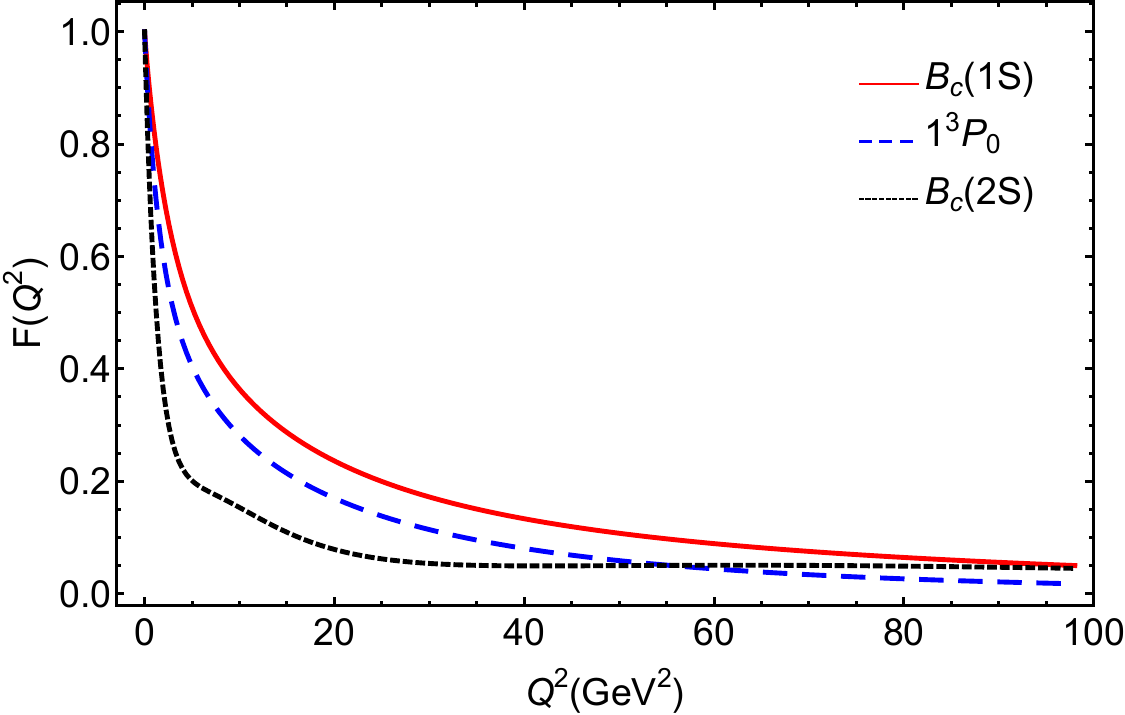}}\\
		\subfloat[charmonium states]{\includegraphics[width=0.32\linewidth]{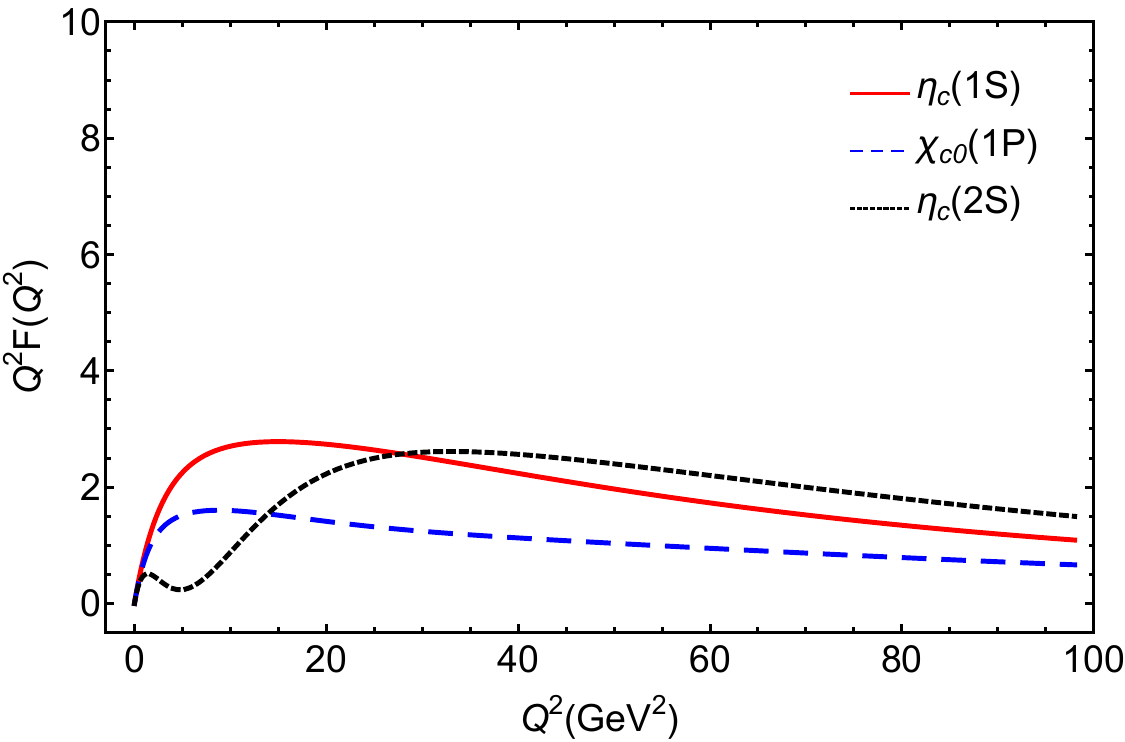}}\quad
		\subfloat[bottomonium states]{\includegraphics[width=0.32\linewidth]{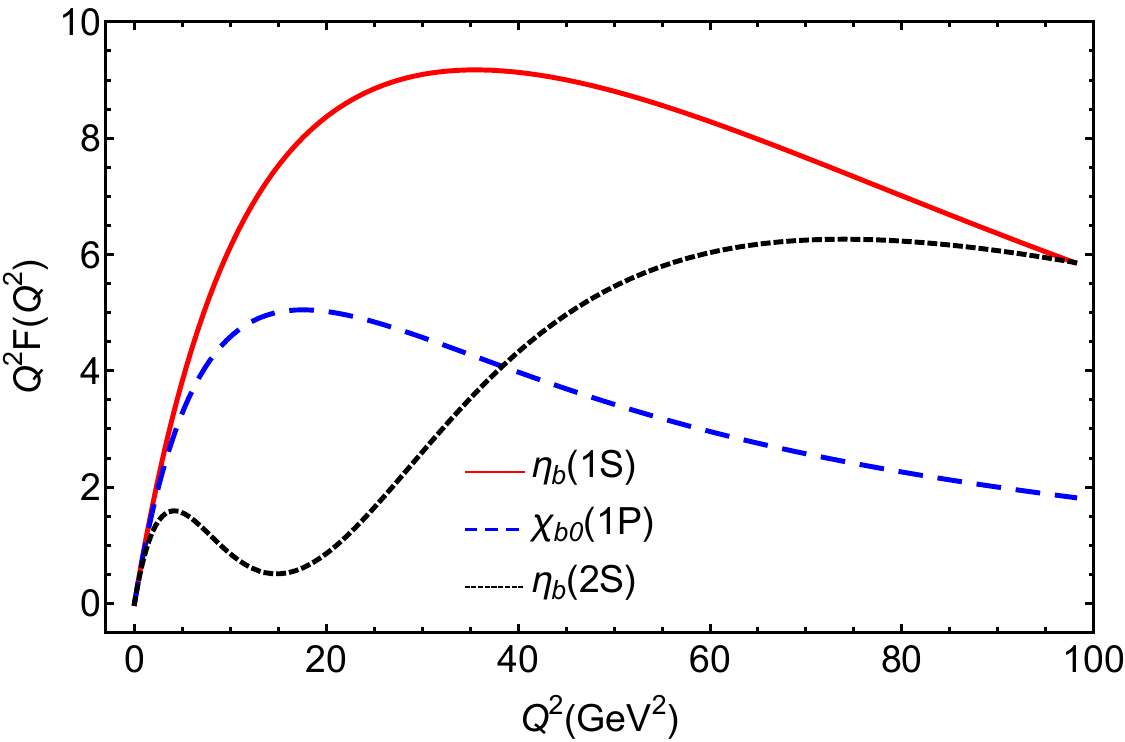}}\quad
		\subfloat[$\rm B_c$ meson states]{\includegraphics[width=0.32\linewidth]{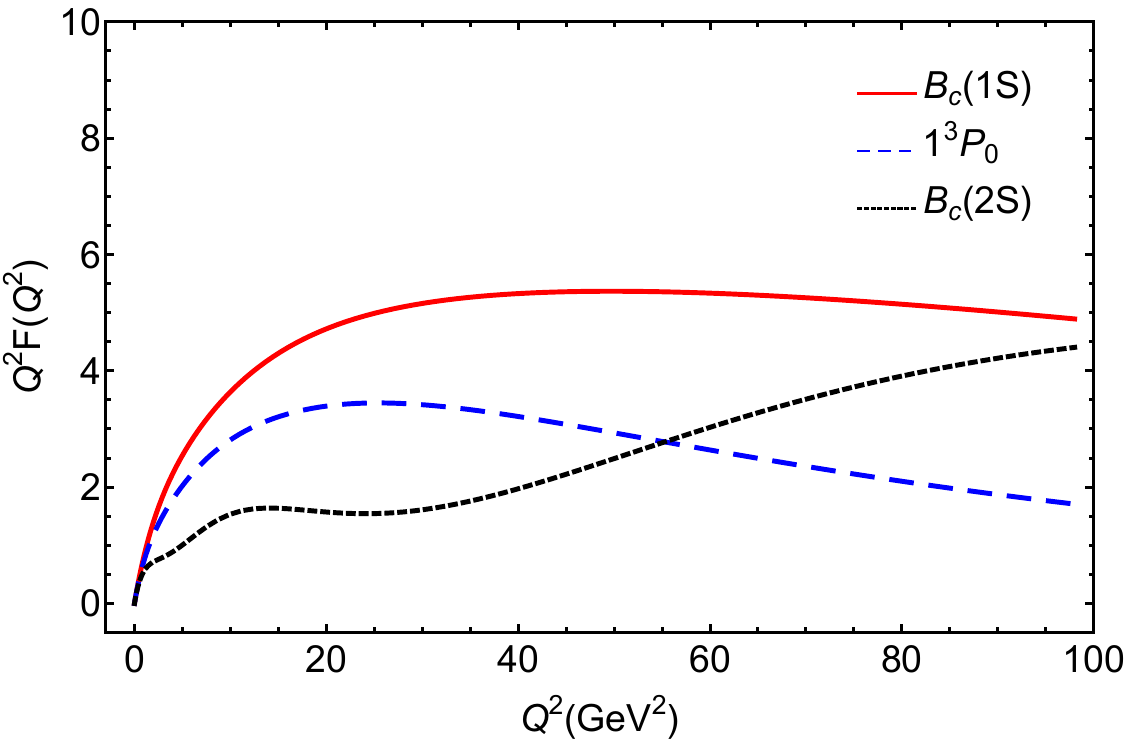}}
		\caption{The electromagnetic FFs and $\rm Q^2$FFs of $J=0$ charmonium, bottomonium and $\rm B_c$ meson states with combined contributions from both $|q\bar{q}\rangle$ and $|q\bar{q}g\rangle$ Fock sectors. The solid, dashed, and dotted lines represent different states. For charmonium and bottomonium states, $F(Q^2)= F^{c(b)}(Q^2)$, see text for details. For $\rm B_c$ meson states, the electromagnetic FFs are weighted by the charge of $b$ and $\bar{c}$ quarks: $F(Q^2)=\frac{1}{3}F^b(Q^2)+\frac{2}{3}F^{\bar{c}}(Q^2)$.}
		\label{ff}
	\end{figure*}
	
	\begin{figure*}
		\centering
		\setcounter{subfigure}{0}
		\subfloat[charmonium states]{\includegraphics[width=0.32\linewidth]{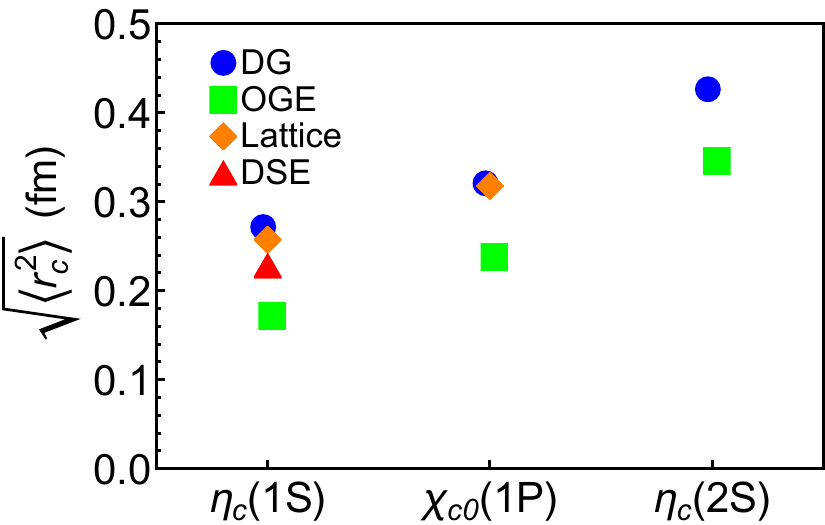}}\quad
		\subfloat[bottomonium states]{\includegraphics[width=0.32\linewidth]{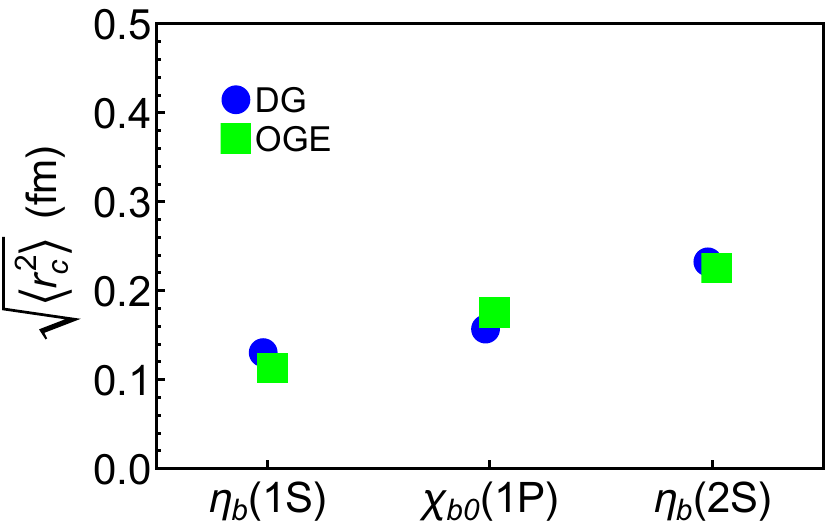}}\quad
		\subfloat[$\rm B_c$ meson states]{\includegraphics[width=0.32\linewidth]{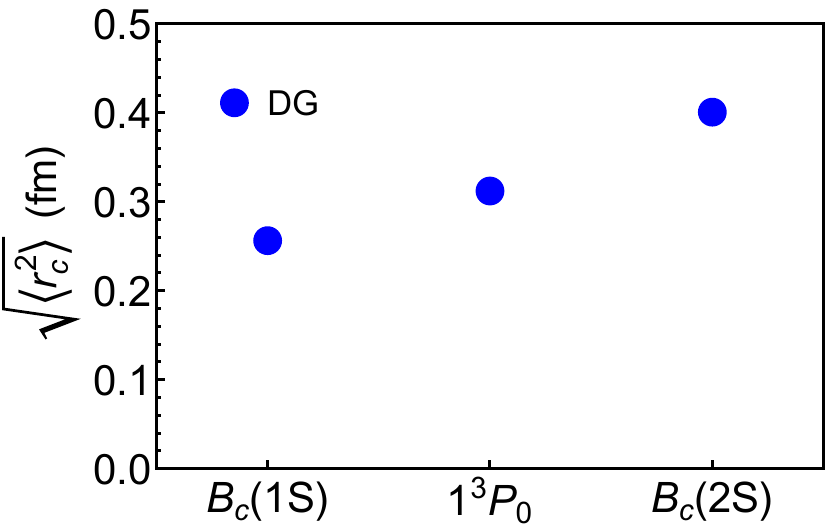}}
		\caption{The charge radii of (pseudo)scalar mesons (solid circles), compared with results from BLFQ OGE (squares)~\cite{y.li1,s.tang}, lattice QCD (rhombuses)~\cite{PhysRevD.73.074507} and DSE (triangles)~\cite{Maris:2006ea}. The BLFQ DG results are obtained from the derivative of the charge form factors (shown in Fig.~\ref{ff}) at zero momentum transfer, see Eq.~\eqref{rc}. The BLFQ OGE results are obtained with $N_{\max}=L_{\max}=32$ for charmonium and bottomonium states. For charmonium and bottomonium states, $\langle r^2_c \rangle$ = $\langle (r^{c(b)}_c)^2 \rangle$. For $\rm B_c$ meson states, the charge radii are weighted by the charge of $b$ and $\bar{c}$ quarks, $\langle r^2_c \rangle$= $\frac{1}{3} \langle (r^b_c)^2 \rangle + \frac{2}{3} \langle (r^{\bar{c}}_c)^2 \rangle$.}
		\label{radii}
	\end{figure*}
	
	In the nonrelativistic limit, the orbital quantum number $L$ and the spin quantum number $S$ can be separately assigned to the eigenstates of the heavy mesons. The spin quantum number $S$ can be identified from the dominant spin configuration, see Table~\ref{norm}. The orbital quantum number $L$ can be inferred from the angular dependence of the LFWFs in the $|q\bar{q}\rangle$ sector: For example, among the first eight states of the heavy mesons, the second through fifth excited states exhibit an azimuthal nodal structure and thus can be identified as P-waves, while the remaining are S-waves. A common excitation pattern is observed for the P-wave charmonium and bottomonium systems: Their second and fourth excited states show excitations in the longitudinal ($x$) directions, whereas the third and fifth states show excitations in the transverse ($p_\perp$) directions. 
	This characteristic pattern observed in the heavy quarkonia provides plausible guidance for identifying the corresponding P-wave states in the $\rm B_c$ system: The third and fourth excited states of the $\rm B_c$ meson are nearly degenerate in mass and share the same $J^P$ quantum numbers, making them difficult to be distinguished based on the spectroscopy alone. However, drawing on the analogy with the charmonium and bottomonium systems, we tentatively assign $\rm 1^3P_1(1^1P_1)$ to the third (fourth) excited state based on their excitations in the longitudinal (transverse) direction. Under this assumption, the ordering of the $\rm 1^3P_1$ and $\rm 1^1P_1$ states in this calculation is inverted relative to the results from the earlier BLFQ results based on the one-gluon exchange effective interaction~\cite{s.tang}. Specifically, the $\rm 1^3P_1$ state is heavier than the $\rm 1^1P_1$ state by around 40 MeV.

	\subsection{Electromagnetic Form Factor and Radii}\label{eff}
	The electromagnetic form factors (FFs) characterize the electric charge and magnetization distribution inside a composite particle. They quantify the deviation of the target's electromagnetic response from that of a point-like particle. The electromagnetic FFs are experimentally probed via elastic scattering. In this work, for simplicity, we study the electromagnetic FFs for the scalar and pseudoscalar mesons with $J=0$. For these spin-0 states, there exists only one electromagnetic FF which can be defined through the matrix element of the ``+" component of the electromagnetic current operator as~\cite{y.li1,pos6},
	\begin{equation}
		F\left(Q^{2}\right) \triangleq \frac{1}{2 P^{+}}\left\langle\Psi^{*}\left(P^{\prime}\right)\left|j^{+}(0)\right| \Psi(P)\right\rangle,
	\end{equation}
	where $Q^2$=$(P'-P)^2$ is the squared momentum transfer in the elastic scattering.
	
	The electromagnetic FFs for $J=0$ mesons can be expressed in terms of the overlap of the LFWFs as,
	\begin{equation}
		\begin{aligned}
			F&\left(Q^2\right)=\sum_{\mathcal{N}, s_i} \int \left[\mathrm{d}^3p\right]_\mathcal{N}\\ 
			&\times\Psi_{\mathcal{N}}^{\left\{s_i\right\} *}\left(\left\{x_i, \vec{p}_{\perp i}^{~\prime}\right\}\right)
			\Psi_{\mathcal{N}}^{\left\{s_i\right\}}\left(\left\{x_i, \vec{p}_{\perp i}\right\}\right).
		\end{aligned}
	\end{equation}
	
	The shifted transverse momenta in the final-state wavefunctions are defined as,
	\[
	\vec{p}_{\perp i}^{~\prime} = 
	\begin{cases}
		\vec{p}_{\perp i} - (1 - x_i)\vec{\Delta}_\perp, & \text{for the struck parton}, \\
		\vec{p}_{\perp i} + x_i\vec{\Delta}_\perp, & \text{for spectator partons},
	\end{cases}
	\]
	where $\Delta=P-P'=(0,0,\vec{\Delta_\perp})$ is the momentum transfer and $\Delta^2=-Q^2$.
	
	From the electric FF, one can calculate the charge radius, which measures the spatial extent of the system's charge distribution. In non-relativistic quantum mechanics, the root-mean-square radius is defined as the expectation value of the squared displacement operator $\langle r^2 \rangle$. However, in quantum field theory, such a definition in coordinate space is ill-defined due to the absence of consistent local position operators. Instead, the charge radii are defined as the slopes of the electromagnetic FFs at zero momentum transfer ($Q^2=0$),
	\begin{equation}\label{rc}
		\begin{aligned}
			\left\langle r_\mathrm{c}^2 \right\rangle = -6\lim_{Q^2\rightarrow0} \frac{\partial}{\partial Q^2}F\left(Q^{2}\right). \\
		\end{aligned}
	\end{equation}
	
	\begin{figure*}
		\centering
		\setcounter{subfigure}{0}
		\subfloat[charmonium states]{\includegraphics[width=0.32\linewidth]{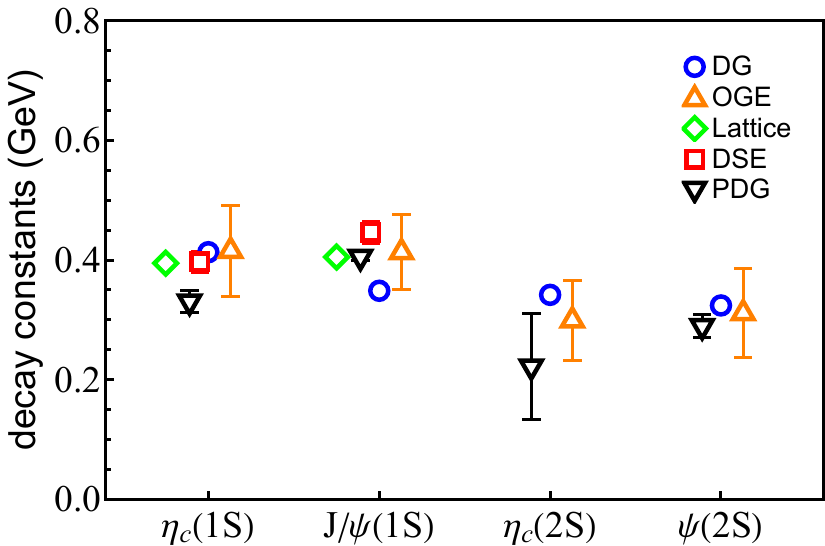}}\quad
		\subfloat[bottomonium states]{\includegraphics[width=0.32\linewidth]{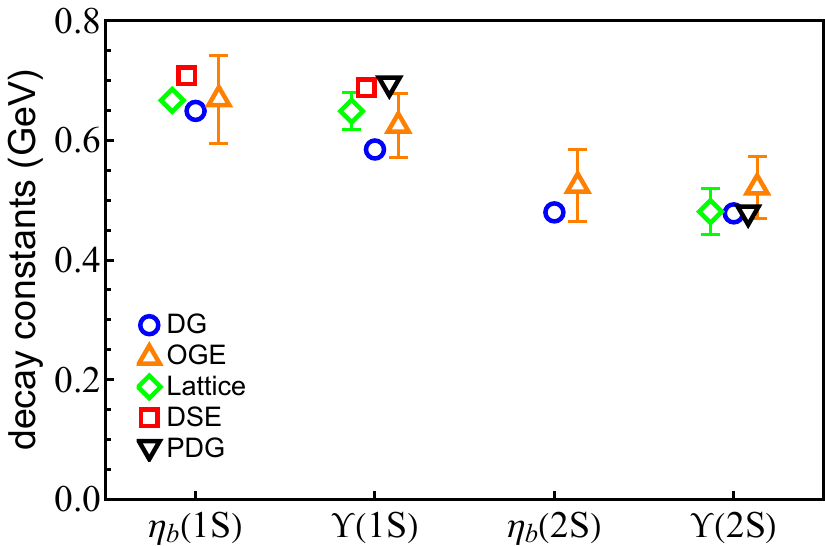}}\quad
		\subfloat[$\rm B_c$ meson states]{\includegraphics[width=0.32\linewidth]{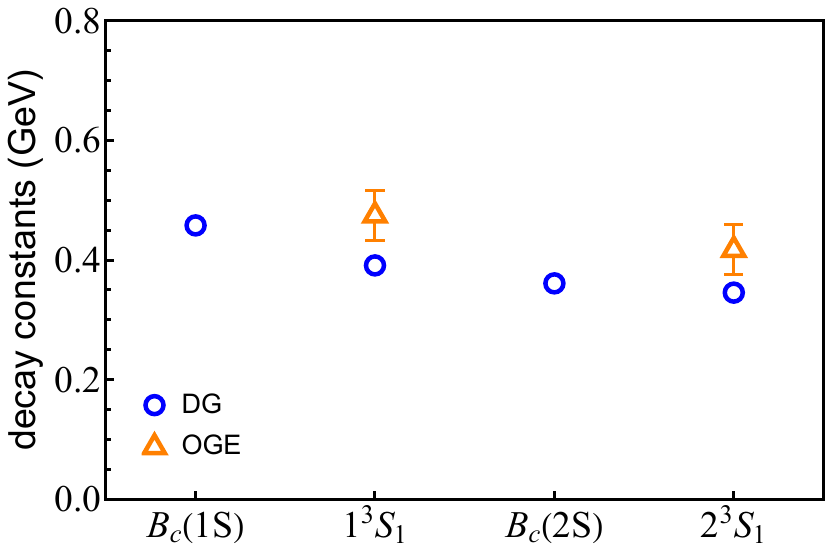}}
		\caption{The decay constants for pseudoscalar and vector states of charmonium and bottomonium, and vector $\rm B_c$ meson states (circles). $M_J=0$ states are used for vector states, see text for details. Results are compared with the BLFQ OGE (regular triangles)~\cite{y.li1,s.tang}, lattice QCD (rhombuses)~\cite{PhysRevD.86.074503,PhysRevD.82.114504,PhysRevD.86.094501,PhysRevD.91.074514}, Dyson-Schwinger Equation (squares)~\cite{PhysRevD.84.096014}, and the experimental data from PDG (inverted triangles)~\cite{pdg}.}
		\label{dcf}
	\end{figure*}
	\begin{figure*}
		\centering
		\setcounter{subfigure}{0}
		\subfloat[charmonium 1S states]{\includegraphics[width=0.32\linewidth]{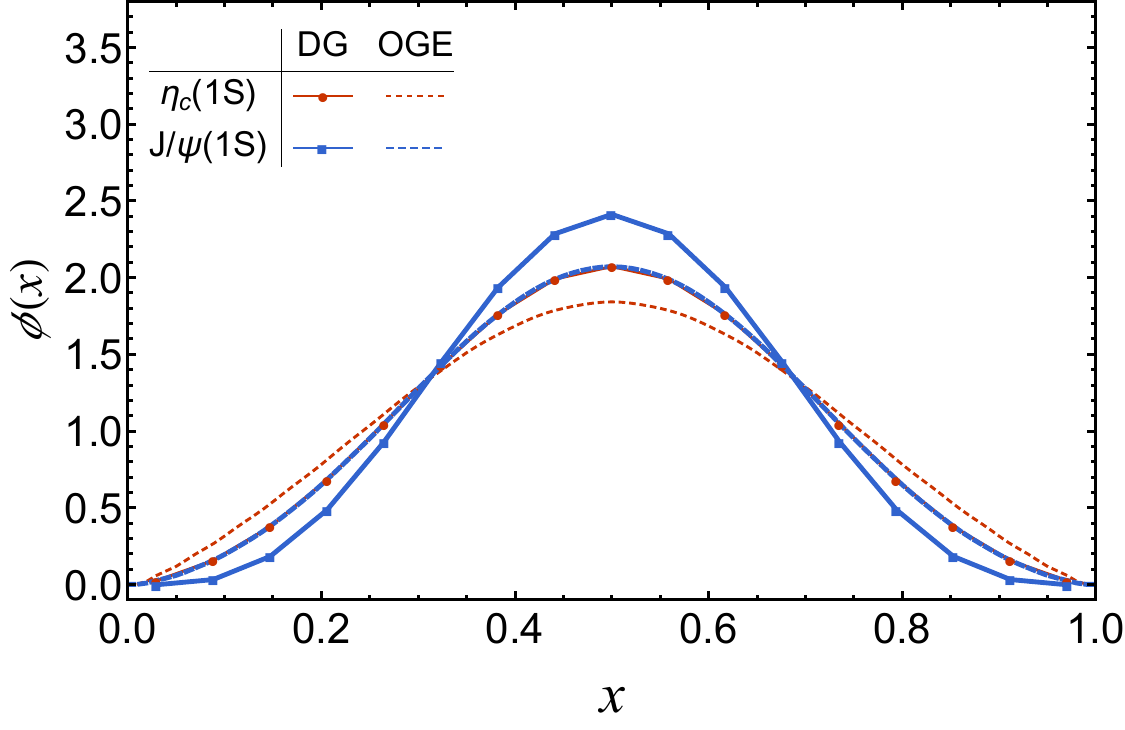}}\quad
		\subfloat[bottomonium 1S states]{\includegraphics[width=0.32\linewidth]{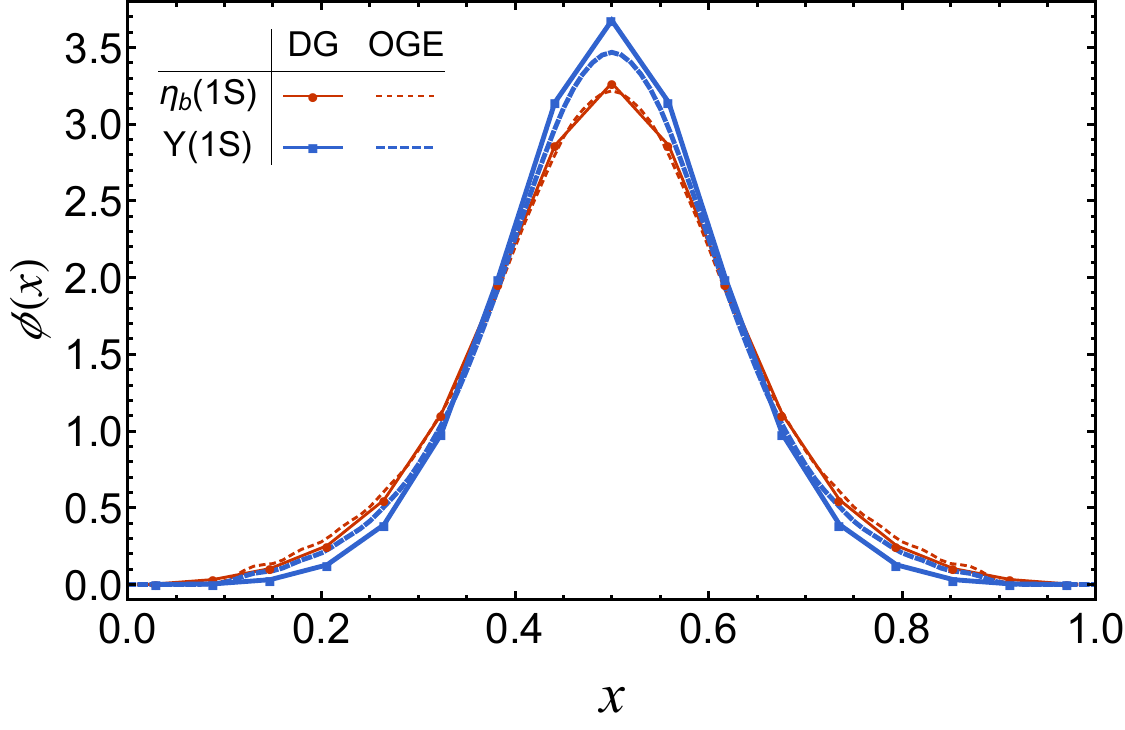}}\quad
		\subfloat[$\rm B_c$ meson 1S states]{\includegraphics[width=0.32\linewidth]{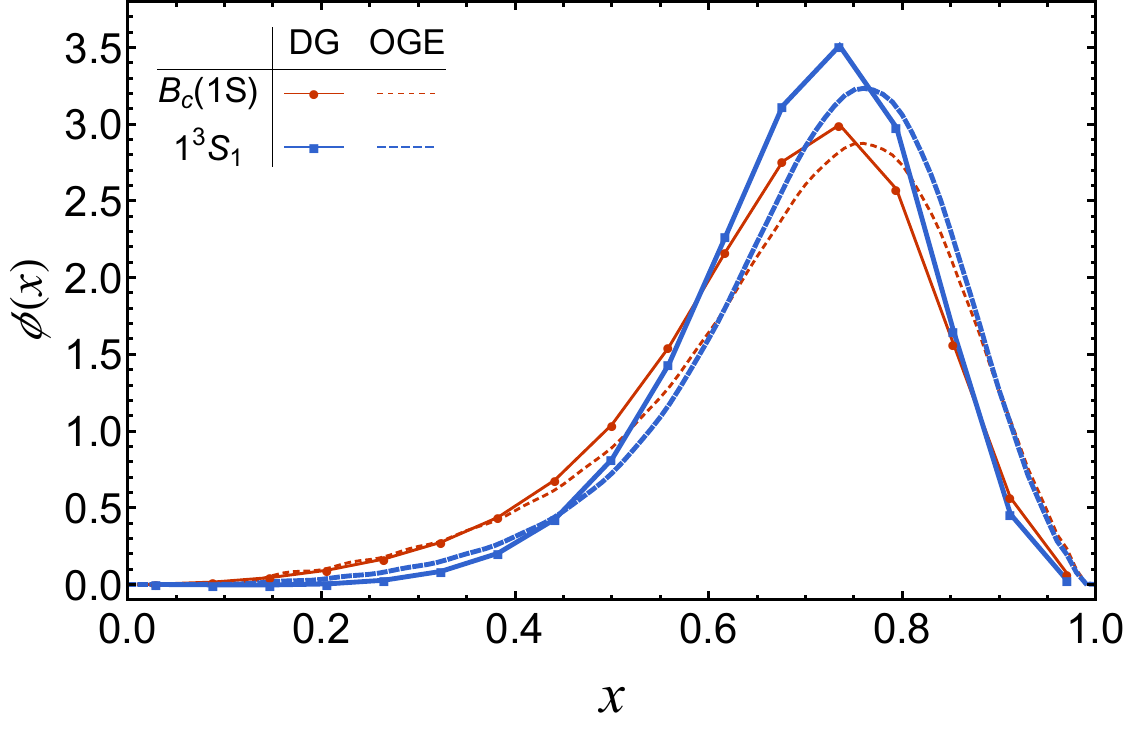}}\\
		\subfloat[charmonium 2S states]{\includegraphics[width=0.32\linewidth]{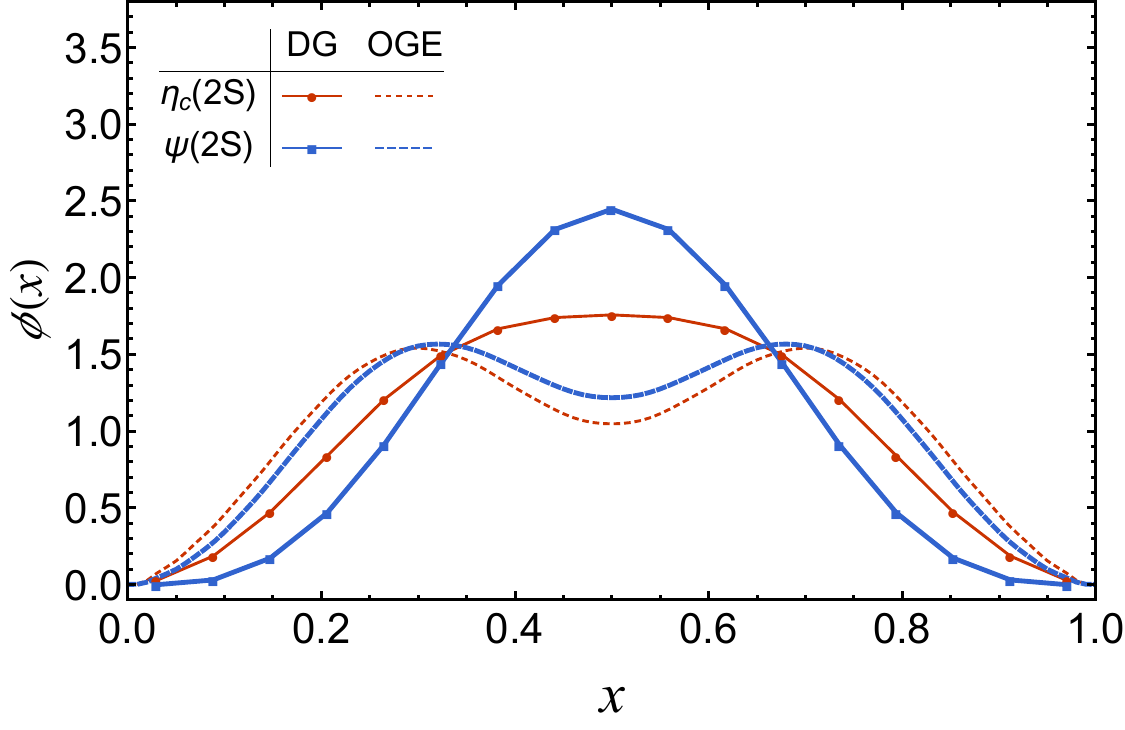}}\quad
		\subfloat[bottomonium 2S states]{\includegraphics[width=0.32\linewidth]{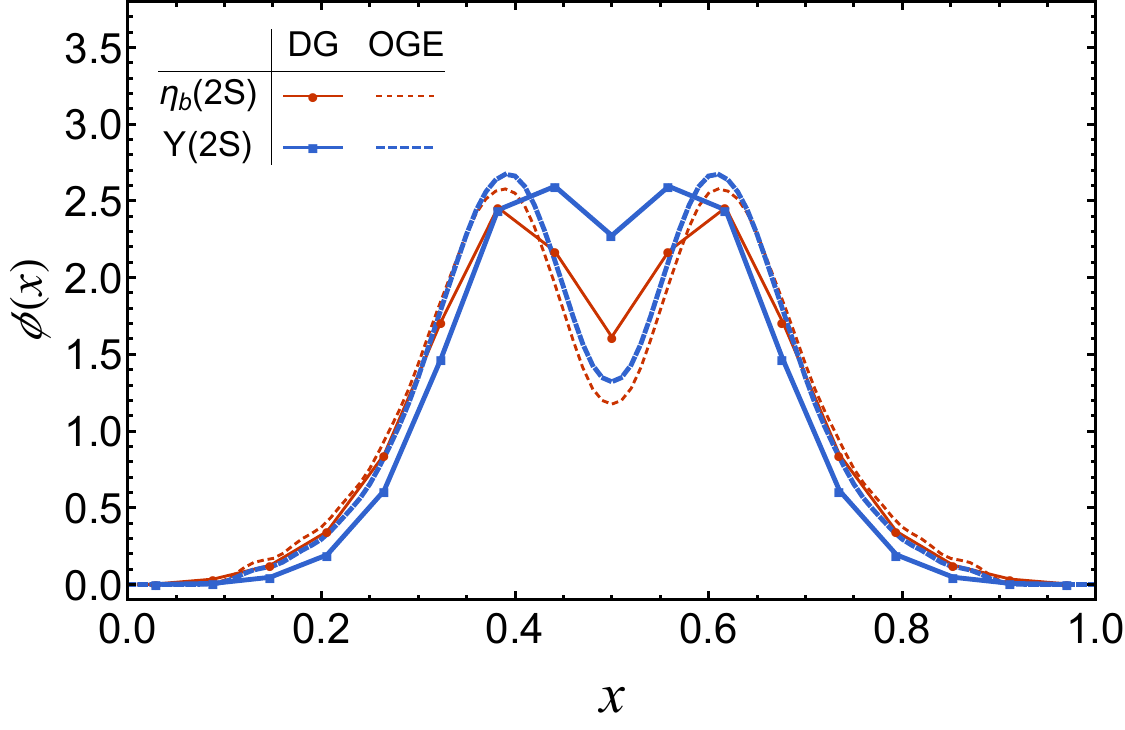}}\quad
		\subfloat[$\rm B_c$ meson 2S states]{\includegraphics[width=0.32\linewidth]{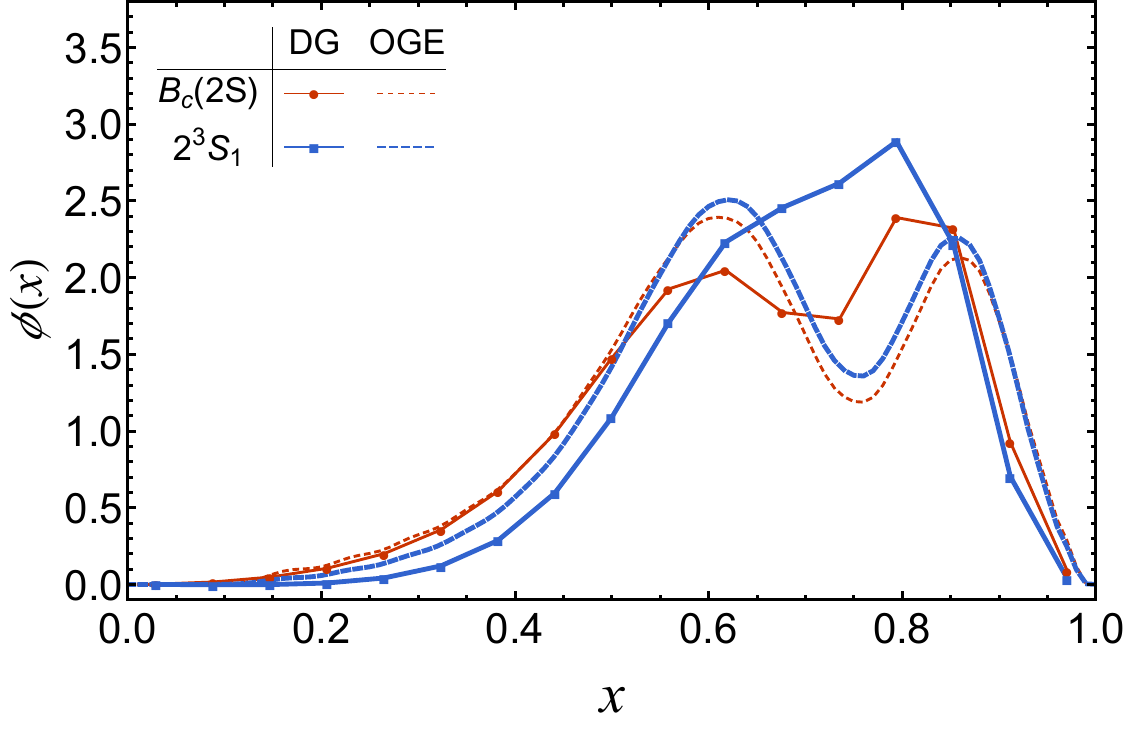}}
		\caption{The quark PDAs of the 1S and 2S charmonium, bottomonium, and $\rm B_c$ meson states (for quark $b$) within the $M_J=0$ sector. The solid and dashed lines represent the DG and OGE result~\cite{y.li1,s.tang} respectively. The blue thick (red thin) lines denote vector (pseudoscalar) states.}
		\label{pda}
	\end{figure*}
	
	In Fig.~\ref{ff}, we present the charge form factors $F(Q^2)$ and $Q^2F(Q^2)$ as functions of momentum transfer, $Q^2$, for the charmonium, bottomonium, and $\rm B_c$ meson states.
	For charmonium and bottomonium systems, the total electromagnetic FFs and charge radii should vanish due to charge conjugation symmetry, so here we consider instead the charge radii of a single quark as the fictitious charge radii~\cite{y.li1}.
	
	The electromagnetic FFs for all three systems decrease with $Q^2$ at small $Q^2$, and approach zero when $Q^2$ is sufficiently large. The steeper slopes at $Q^2=0$ for charmonium states show that charmonium states have larger electromagnetic radii than bottomonium states, which suggests that the bottomonium states are more tightly bound compared to the corresponding charmonium states. Compared to the ground states, the excited states show steeper slopes at $Q^2=0$, which translates to larger charge radii for excited states. This trend is in agreement with the nonrelativistic picture, that is, the excited states are more loosely bound compared to the ground states. For $\eta_c(\mathrm{2S})$ and $\eta_b(\mathrm{2S})$, their FFs show a local minimum at $Q^2\sim5(15) \mathrm{GeV^2}$, which originates from the radial nodal structure of their LFWFs, cf.~Fig.~\ref{wave}.
	For the charmonium systems and the ground state of the bottomonium and $\rm B_c$ system, $Q^2 F(Q^2)$ initially rises and then exhibits a slow decrease in the high-$Q^2$ region. This trend signals the onset of the scaling behavior expected from perturbative QCD, that is, $F(Q^2) \sim 1/Q^2$ at large momentum transfer $Q^2$. The decreasing trend at large $Q^2$ originates from the Gaussian tail of the 2D-HO basis functions adopted in this work.
	Compared to the charmonium systems, the transition to the perturbative regime happens at a larger $Q^2$ in the bottomonium and $\rm B_c$ systems due to their larger system mass. For the 2S states of bottomonium and $\rm B_c$ systems, the scaled form factors continue to rise with $Q^2$ up to 60 (100) $\rm GeV^2$ for bottomonium ($\rm B_c$) states, which suggests a much larger transition scale for the excited states compared to the ground states as a result of the radial nodal structure in their LFWFs.

	In Fig.~\ref{radii} we compare our root-mean-square charge radii of scalar and pseudoscalar mesons with the results from the BLFQ OGE~\cite{y.li1,s.tang}, lattice QCD~\cite{PhysRevD.73.074507}, and Dyson-Schwinger Equation approach~\cite{Maris:2006ea}. 
	The electromagnetic radii exhibit an increasing trend with excitation energy. Compared to the BLFQ OGE calculations, our charmonium results are systematically larger and closer to the lattice QCD results.
	We attribute this difference to the contribution from the $|q\bar{q}g\rangle$ Fock sector. The absence of a confining potential in the $|q\bar{q}g\rangle$ sector leads to a spatially extended distribution for this Fock component, thereby inflating its contribution to the total radius. This effect is more pronounced for the charmonium states due to their comparatively larger contribution from the $|q\bar{q}g\rangle$ sector.
	We expect that a more realistic estimate of the size of the $|q\bar{q}g\rangle$ component will be achieved as higher Fock sectors are progressively included in our basis, so that the confinement in the higher Fock sectors can dynamically emerge.
	
	\subsection{Decay Constants}
	Decay constants encapsulate the nonperturbative dynamics required to calculate the leptonic decay rates. Physically, the pseudoscalar decay constant $f_P$ describes the weak annihilation process mediated by a $W$ boson, while the vector decay constant $f_V$ describes the process mediated by a virtual photon. The decay constants $f_{P,V}$ are defined as the local vacuum-to-hadron transition matrix elements of the electroweak currents. They characterize the amplitude for the quark-antiquark pair to annihilate into the vacuum through the electroweak currents~\cite{y.li1,pdg},
	\begin{equation}
		\begin{array}{c}
			\left\langle 0\left|\bar{\Psi}(0) \gamma^{+} \gamma_{5} \Psi(0)\right| \Psi(P)\right\rangle=P^{+} f_{P}, \\
			\left\langle 0\left|\bar{\Psi}(0) \gamma^{+} \Psi(0)\right| \Psi(P)\right\rangle=P^{+} f_{V}.
		\end{array}
	\end{equation}
	
	Here we employ the ``good current" component of the electroweak current, $J^+_5 = \bar{\Psi}\gamma^+\gamma_5\Psi$ for pseudoscalar mesons and $J^+ = \bar{\Psi}\gamma^+\Psi$ for vector mesons. This choice, together with choosing $M_J=0$ component for vector states, allows us to express the matrix elements using only the dynamical components of the fermion fields~\cite{m.li1}.
	Due to the triviality of the light-front vacuum, the matrix elements for leptonic decays are free from vacuum fluctuation contributions, which means that the decay constants only receive contributions from the leading Fock sector $|q\bar{q}\rangle$ in the heavy meson LFWFs~\cite{BRODSKY1998299}.
	Using the ``good" component of the current and $M_J=0$ component of the heavy meson states, the decay constants $f_{P,V}$ are expressed in terms of the three-dimensional momentum integration of the LFWFs in the $|q\bar{q}\rangle$ sector as~\cite{PhysRevD.22.2157,y.li1},
	\begin{equation}
		\frac{f_{P,V}}{2 \sqrt{2 N_{C}}}=\int\left[\mathrm{d}^{3} p\right]_{2} \Psi_{2}^{\downarrow\uparrow\mp\uparrow\downarrow}\left(\{x_i,\vec{p}_{\perp i}\}\right),
	\end{equation}
	where $\Psi_2$ is the leading Fock sector LFWF of the heavy meson states with $M_J=0$,
	and the plus (minus) sign is associated with the vector (pseudoscalar) state. Physically, the decay constants correspond to the wave function at the origin in the coordinate space in the $|q\bar{q}\rangle$ sector.
	
	In Fig.~\ref{dcf}, we compare our decay constants of pseudoscalar and vector mesons with the results from the BLFQ OGE~\cite{y.li1,s.tang}, lattice QCD~\cite{PhysRevD.86.074503,PhysRevD.82.114504,PhysRevD.86.094501,PhysRevD.91.074514}, and Dyson-Schwinger Equation approach~\cite{PhysRevD.84.096014}.
	Our results are generally consistent with the experimental data from the PDG and the previous BLFQ calculations based on the OGE effective interactions. However, compared with the results from lattice QCD and DSE, our values for the vector ground states, $J/\psi(\mathrm{1S})$ and $\Upsilon(\mathrm{1S})$, are systematically smaller. In particular, within the charmonium sector, we observe that the decay constant of $J/\psi(1S)$ is smaller than that of $\eta_c(1S)$, a discrepancy suggesting that including only the leading two Fock sectors might not be enough to properly describe the pseudoscalar and vector decay constants.
	It would be interesting to see whether the relative size of the decay constants for these two states gets reversed as higher Fock sectors are included in the basis.
	Compared to the ground states, the decay constants for the excited states are smaller, which can be understood from the overlap of the LFWFs at the origin in coordinate space: Higher excited states exhibit a more extended spatial distribution, which naturally leads to a reduced value at the origin. 
	In addition, the contribution to the LFWFs from the leading Fock sector, $N_{q\bar{q}}^2$, is smaller in the excited states compared to the ground state, cf.~Table~\ref{norm}, which further reduces the decay constants for the excited states.
	
	\subsection{Parton Distribution Amplitudes}
	The Parton Distribution Amplitudes (PDAs) encode the nonperturbative structural information of partons inside a hadron at the amplitude level. They are often used to describe the exclusive processes at large momentum transfer~\cite{PhysRevD.22.2157,y.li1}.
	Similar to the decay constants, the leading-twist PDAs are defined from the vacuum-to-meson matrix elements~\cite{PhysRevD.22.2157,PhysRevD.74.114028}. The PDAs for the S-wave heavy meson states can be written as the integral of LFWFs over the transverse momentum in the leading Fock sector,
	
	\begin{multline}
		\Phi_{P,V}(x)=\frac{2 \sqrt{2 N_{C}}}{f_{P,V}}\int\left[\mathrm{d}^{3} p\right]_2 \delta(x-x_1)\\
		\times \Psi_{2}^{\downarrow\uparrow\mp\uparrow\downarrow}\left(\{x_i,\vec{p}_{\perp i}\}\right).
	\end{multline}
	Here we adopt the normalization convention $\int_0^1 dx \Phi_{P, V} (x) =1$. $\Psi_2$ is the leading Fock sector LFWF, and the plus (minus) sign is associated with the vector (pseudoscalar) states.
	
	We present the PDAs for the S-wave heavy meson states in Fig.~\ref{pda}. 
	The PDAs for the bottomonium states are significantly narrower and higher than those for the charmonium states. This is a direct consequence of their larger constituent mass, indicating that the bottomonium system is closer to the non-relativistic limit compared to the charmonium system. For the unequal-mass $\rm B_c$ system, the distributions are asymmetric and the PDA for the bottom quark peaks in the large-$x_b$ region, where $x_b$ denotes the longitudinal momentum fraction of the bottom quark. This reflects the kinematics in the asymmetric heavy meson systems where the heavier quark carries the larger part of the meson's longitudinal momentum. Regarding the excited states $\eta_c(2S)$ and $\psi(2S)$, our results exhibit a single-peak profile, unlike the results based on the OGE effective interaction, which display a double-peak structure originating from the radial excitations. This behavior can be traced back to the wave functions shown in Fig.~\ref{wave}, where the nodal structures along the longitudinal direction are weaker compared to those in the transverse directions and those from the OGE effective interaction~\cite{y.li1}. The difference in nodal structures between the longitudinal and transverse directions reflects the dynamical nature of the rotational symmetry on the light front. Studying the behavior of the PDAs as higher Fock sectors are progressively included in the basis will provide important insights into the rotational symmetry of nonperturbative dynamics on the light front.
	
	\begin{figure*}
		\centering
		\setcounter{subfigure}{0}
		\subfloat[$\eta_c(\mathrm{1S})$]{\includegraphics[width=0.36\linewidth]{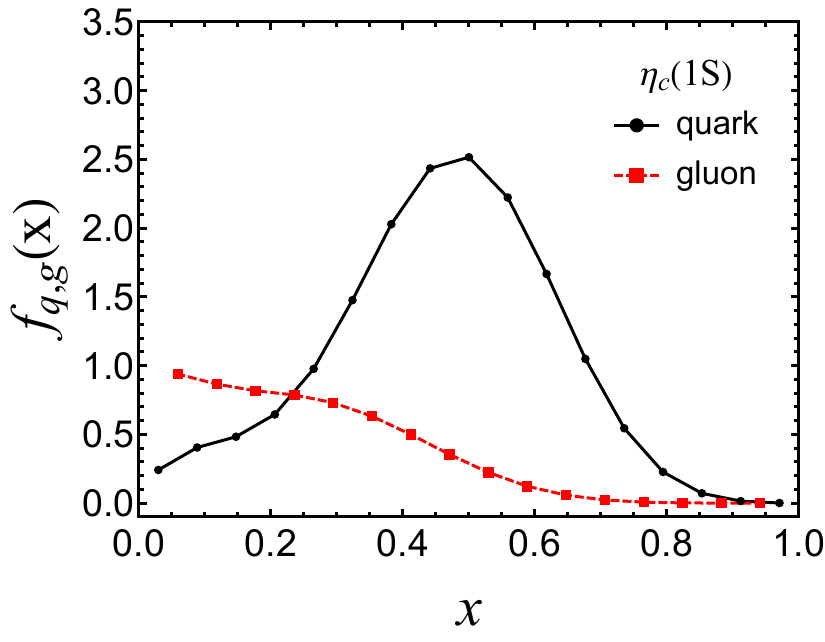}}\quad
		\subfloat[$J/\psi(\mathrm{1S})$]{\includegraphics[width=0.36\linewidth]{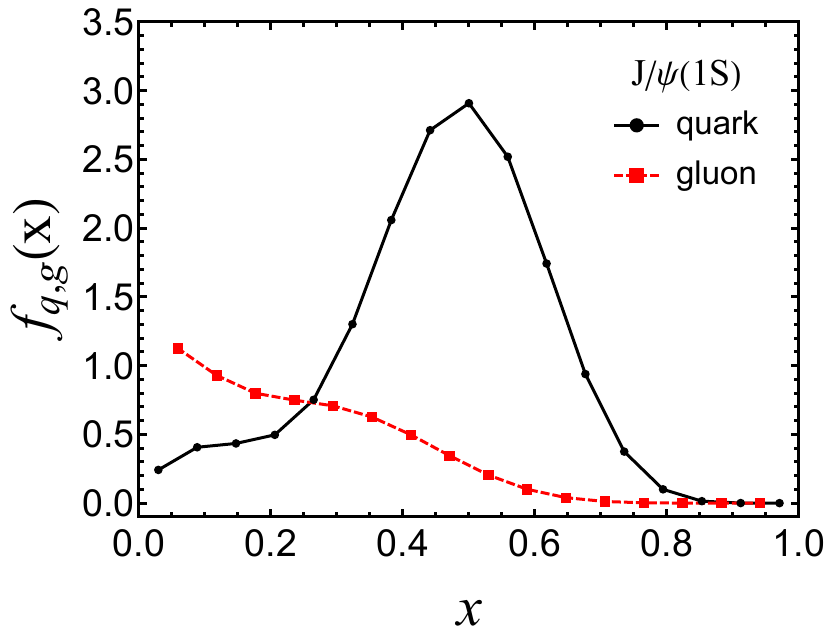}}\quad
		\subfloat[$\chi_{c0}(\mathrm{1P})$]{\includegraphics[width=0.36\linewidth]{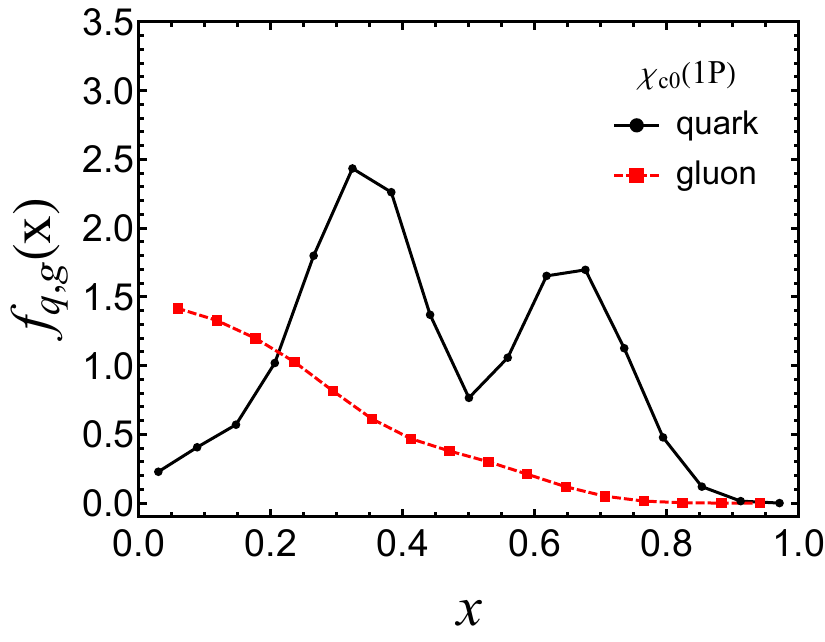}}\quad
		\subfloat[$\chi_{c1}(\mathrm{1P})$]{\includegraphics[width=0.36\linewidth]{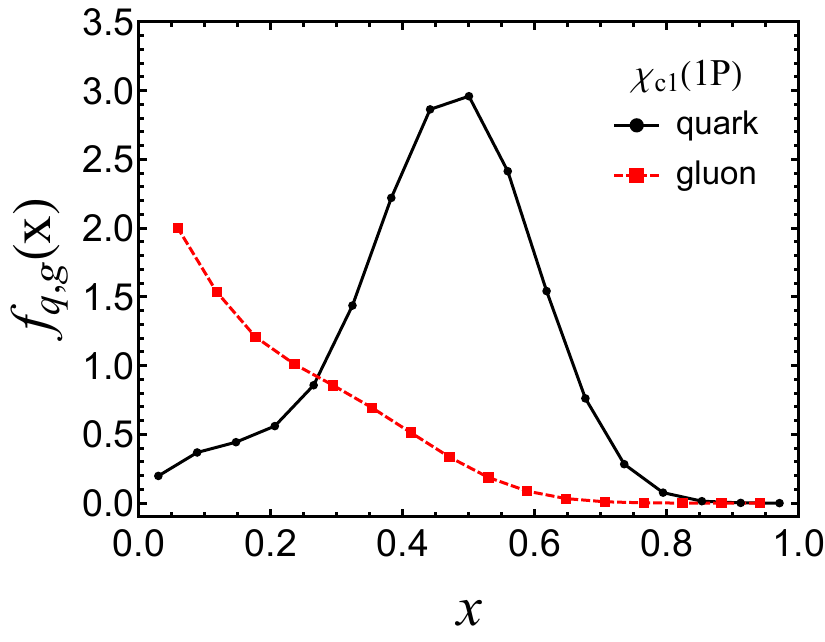}}\quad
		\subfloat[$h_c(\mathrm{1P})$]{\includegraphics[width=0.36\linewidth]{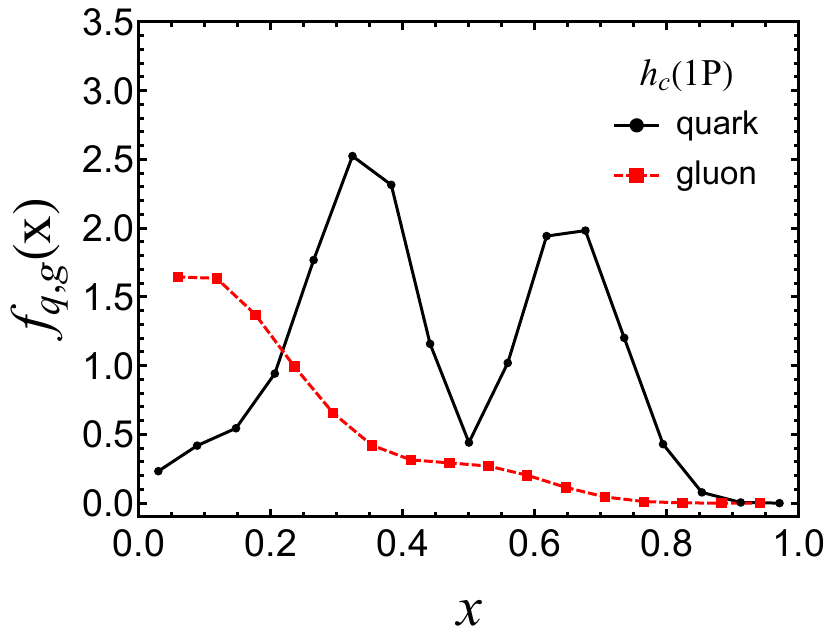}}\quad
		\subfloat[$\chi_{c2}(\mathrm{1P})$]{\includegraphics[width=0.36\linewidth]{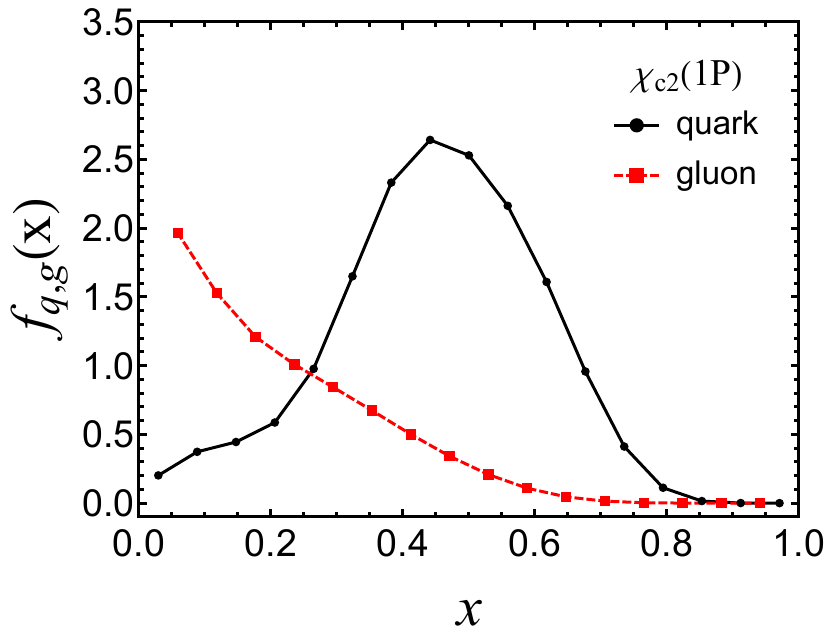}}\quad
		\subfloat[$\eta_c(\mathrm{2S})$]{\includegraphics[width=0.36\linewidth]{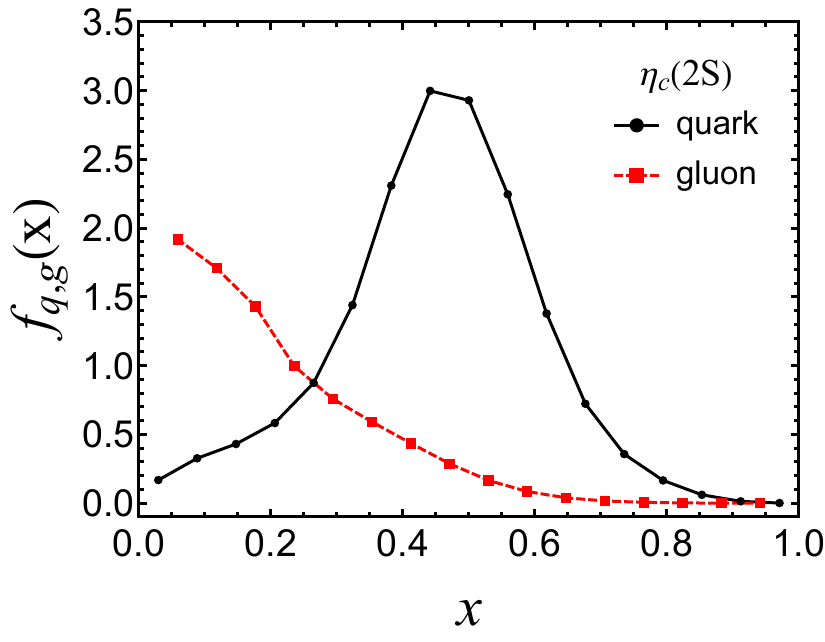}}\quad
		\subfloat[$\psi(\mathrm{2S})$]{\includegraphics[width=0.36\linewidth]{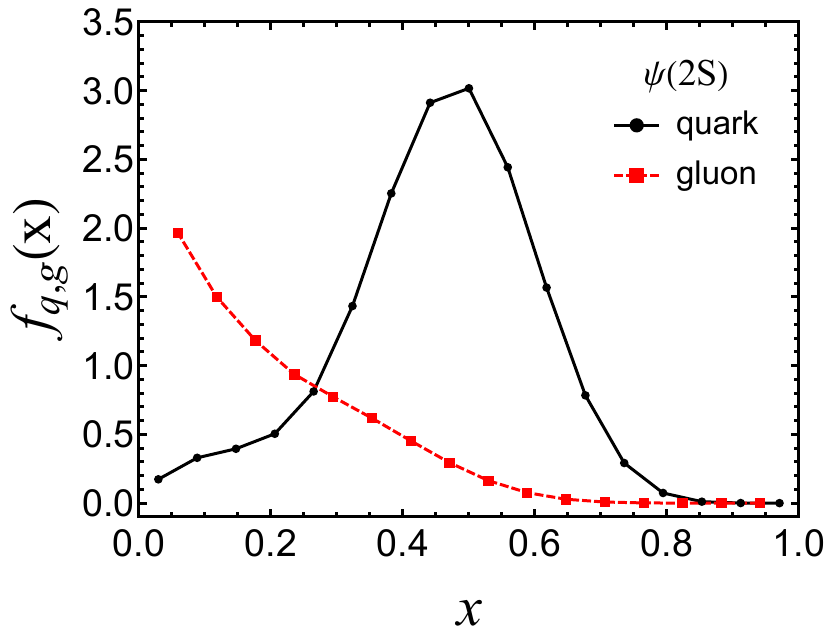}}\\
	\end{figure*}
	\begin{figure*}
		\centering
		\setcounter{subfigure}{8}
		\subfloat[$\eta_b(\mathrm{1S})$]{\includegraphics[width=0.36\linewidth]{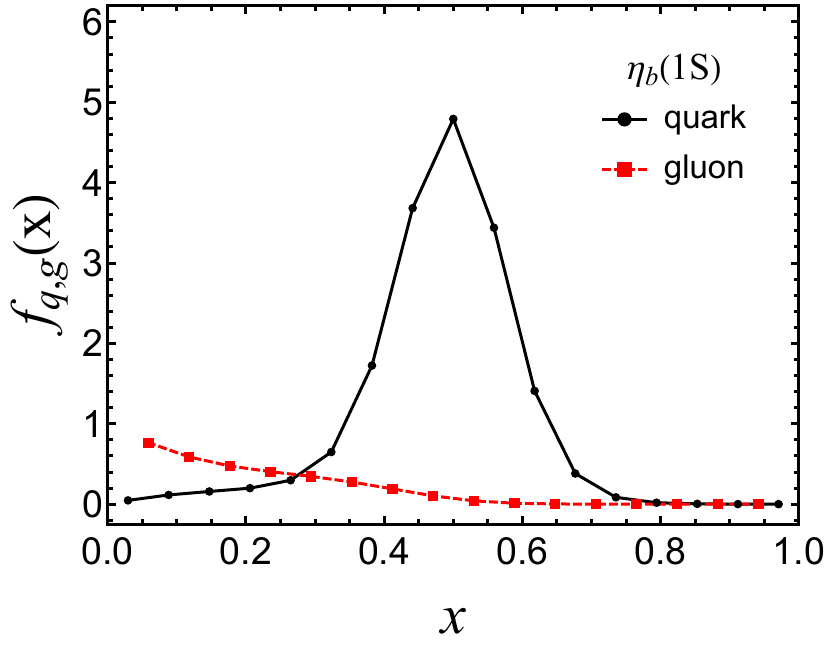}}\quad
		\subfloat[$\Upsilon(\mathrm{1S})$]{\includegraphics[width=0.36\linewidth]{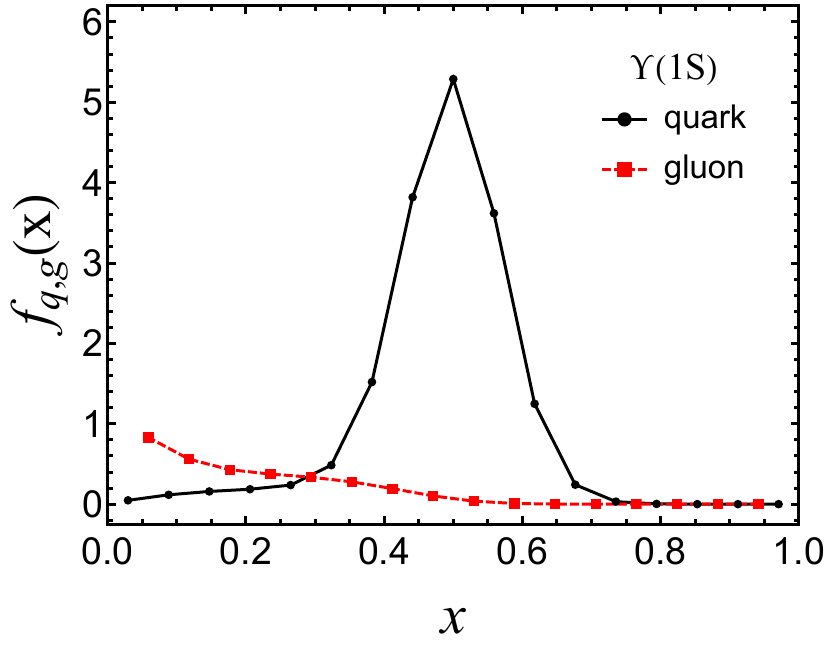}}\quad
		\subfloat[$\chi_{b0}(\mathrm{1P})$]{\includegraphics[width=0.36\linewidth]{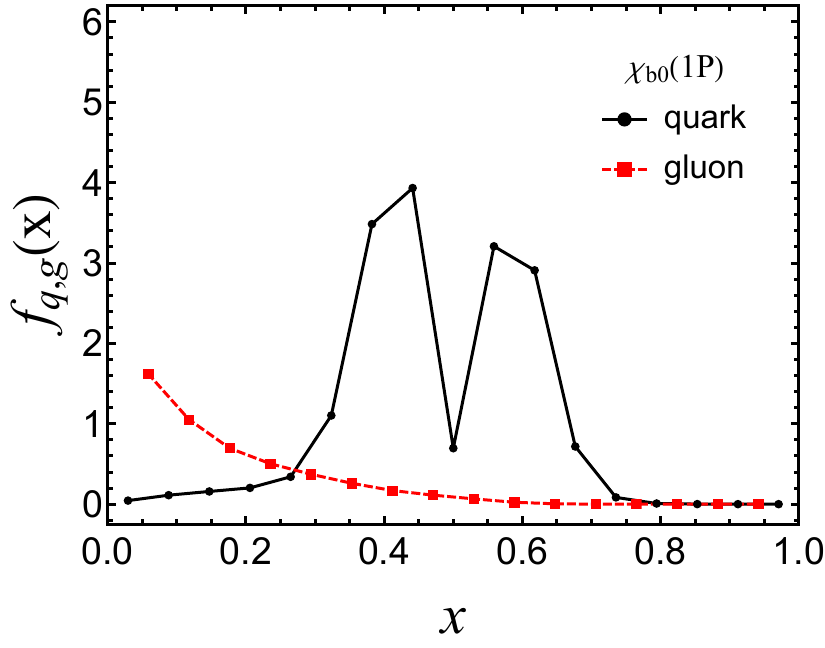}}\quad
		\subfloat[$\chi_{b1}(\mathrm{1P})$]{\includegraphics[width=0.36\linewidth]{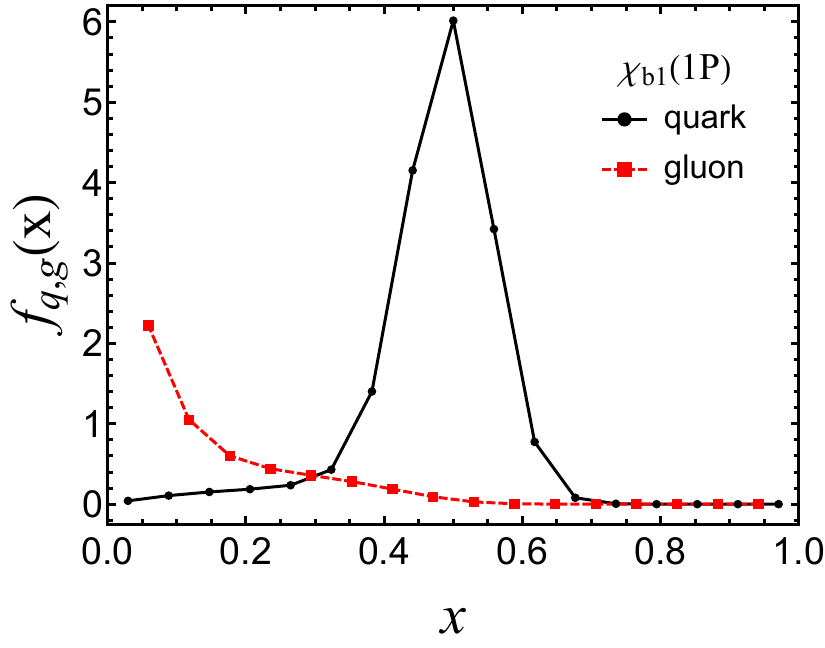}}\quad
		\subfloat[$h_b(\mathrm{1P})$]{\includegraphics[width=0.36\linewidth]{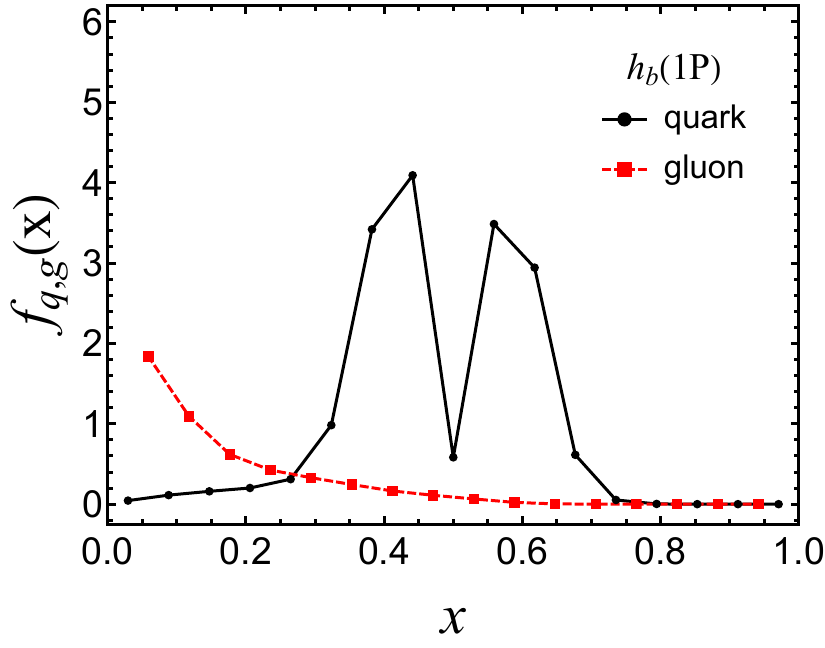}}\quad
		\subfloat[$\chi_{b2}(\mathrm{1P})$]{\includegraphics[width=0.36\linewidth]{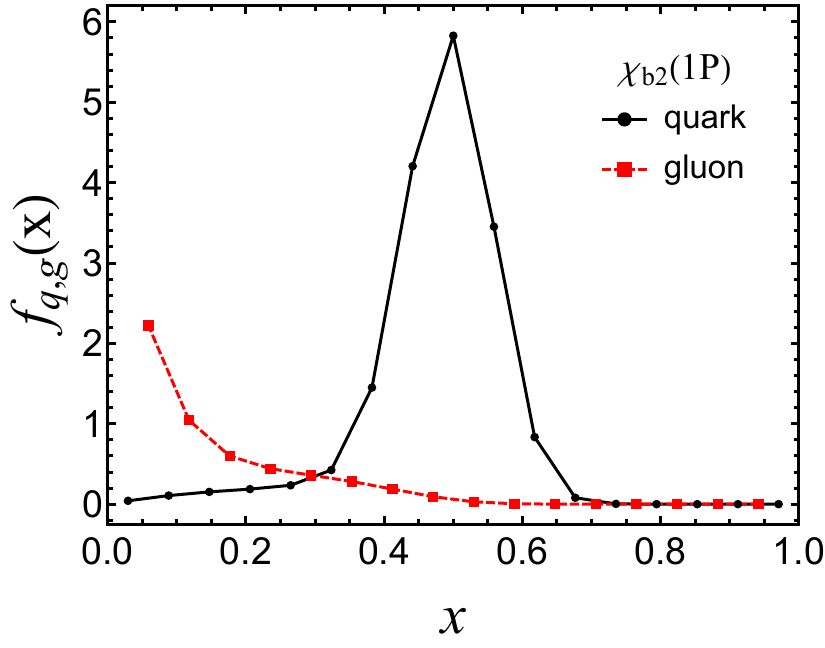}}\quad
		\subfloat[$\eta_b(\mathrm{2S})$]{\includegraphics[width=0.36\linewidth]{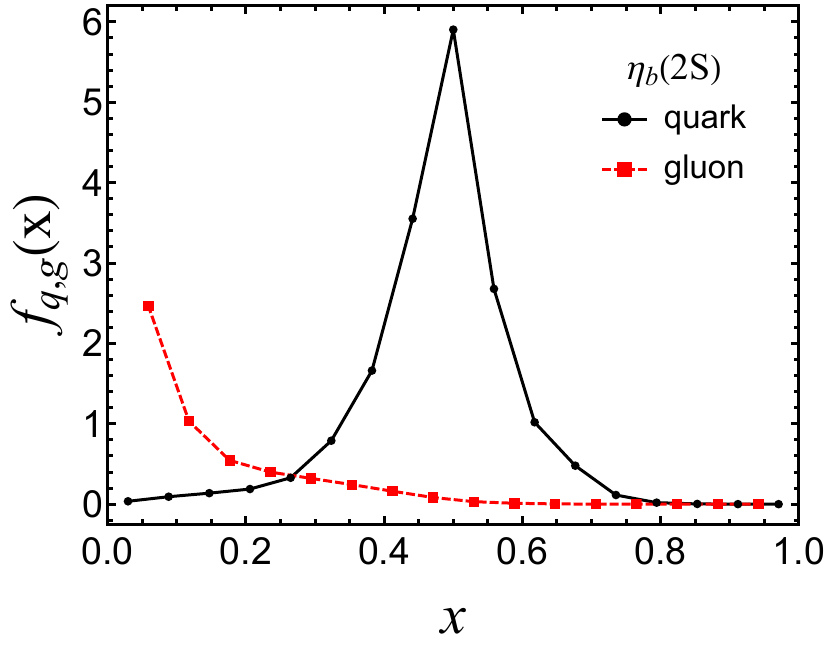}}\quad
		\subfloat[$\Upsilon(\mathrm{2S})$]{\includegraphics[width=0.36\linewidth]{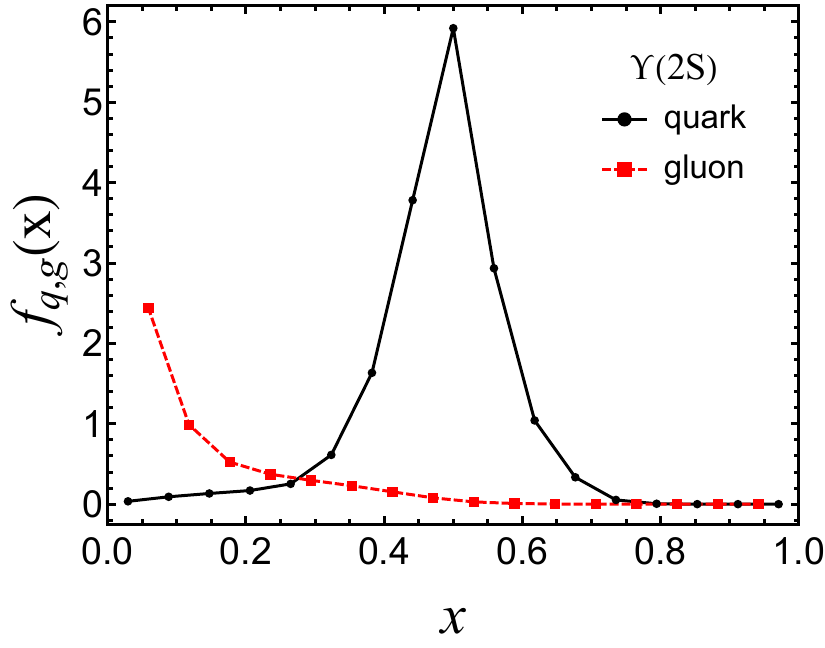}}\\
	\end{figure*}
	\begin{figure*}
		\centering
		\setcounter{subfigure}{16}
		\subfloat[$\rm B_c(\mathrm{1S})$]{\includegraphics[width=0.36\linewidth]{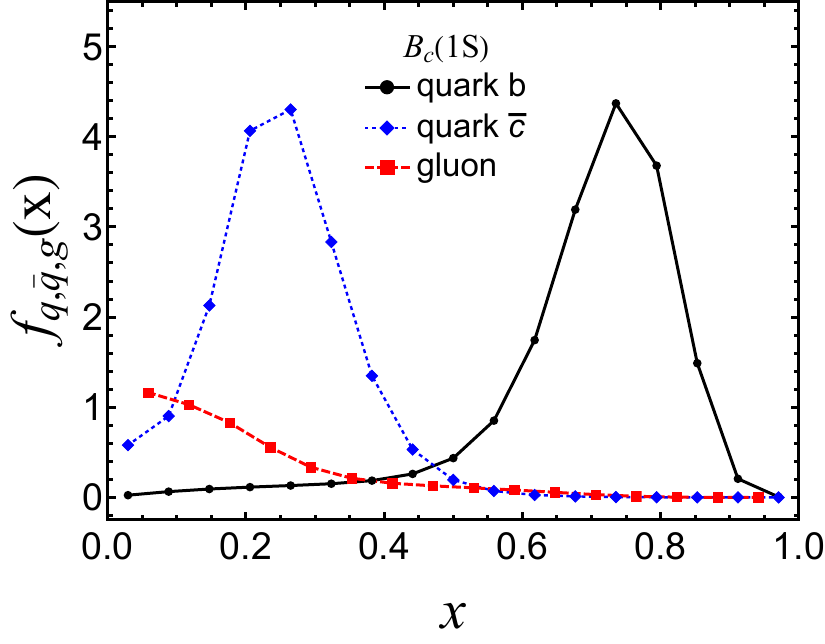}}\quad
		\subfloat[$1^3\mathrm{S_1}$]{\includegraphics[width=0.36\linewidth]{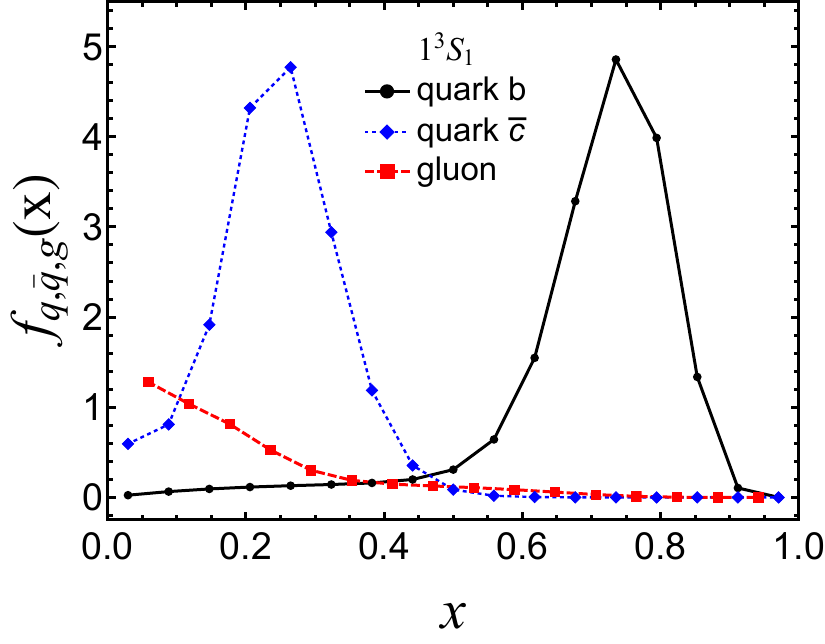}}\quad
		\subfloat[$1^3\mathrm{P_0}$]{\includegraphics[width=0.36\linewidth]{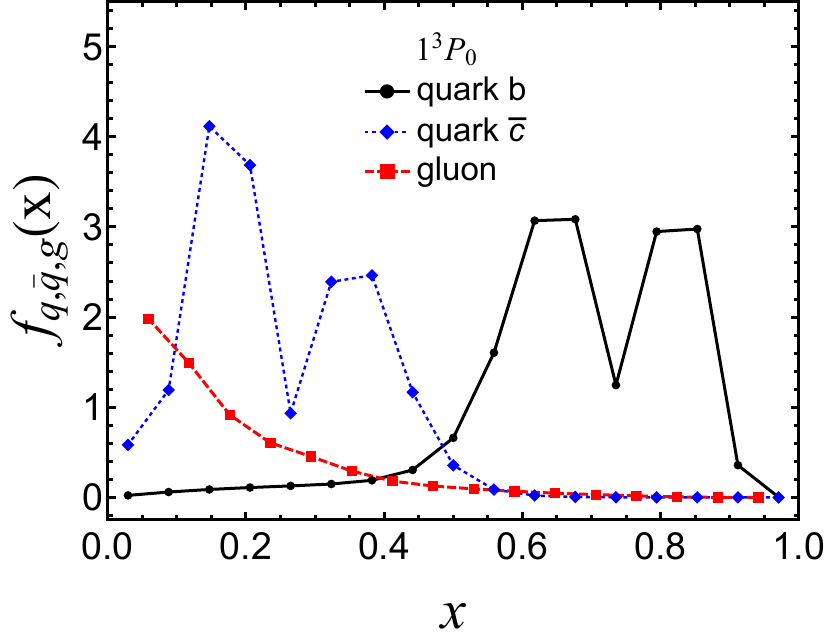}}\quad
		\subfloat[$1^3\mathrm{P_1}$]{\includegraphics[width=0.36\linewidth]{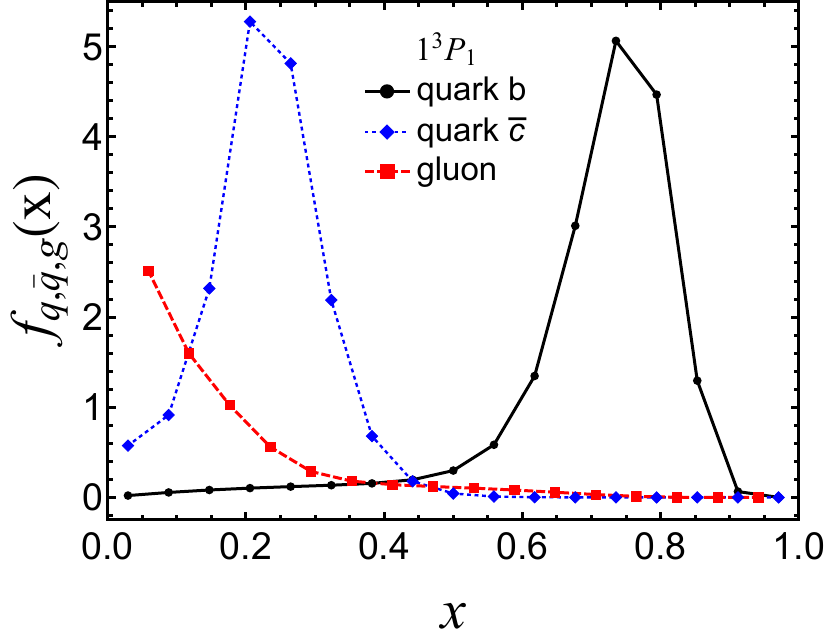}}\quad
		\subfloat[$1^1\mathrm{P_1}$]{\includegraphics[width=0.36\linewidth]{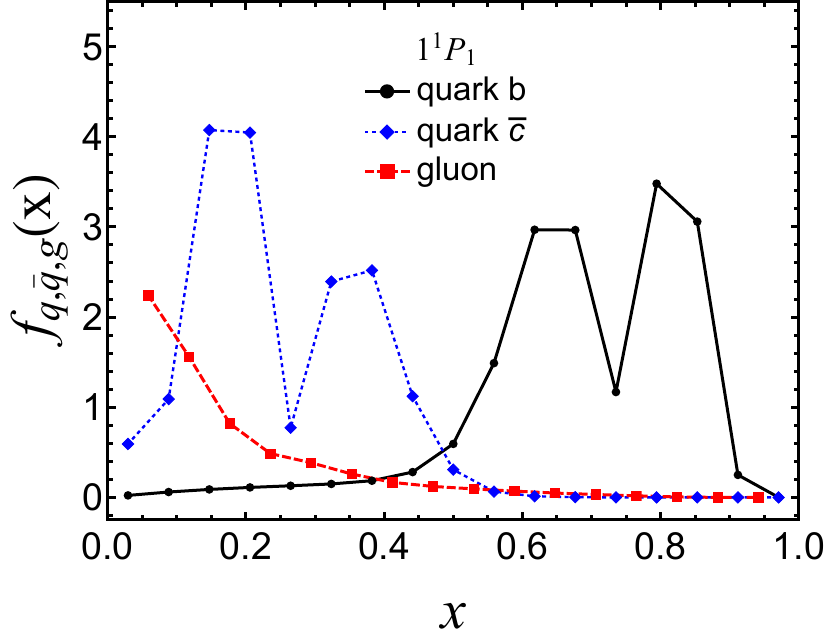}}\quad
		\subfloat[$1^3\mathrm{P_2}$]{\includegraphics[width=0.36\linewidth]{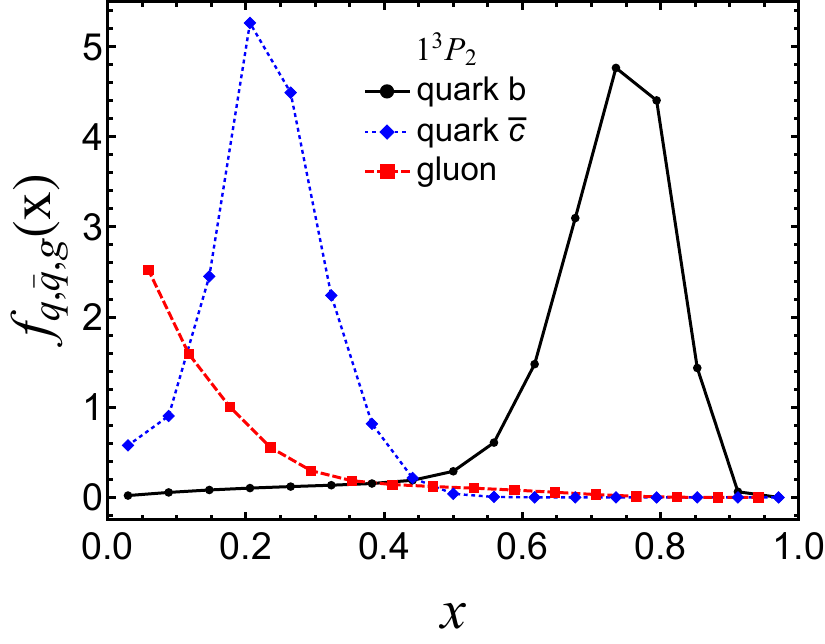}}\quad
		\subfloat[$\rm B_c(\mathrm{2S})$]{\includegraphics[width=0.36\linewidth]{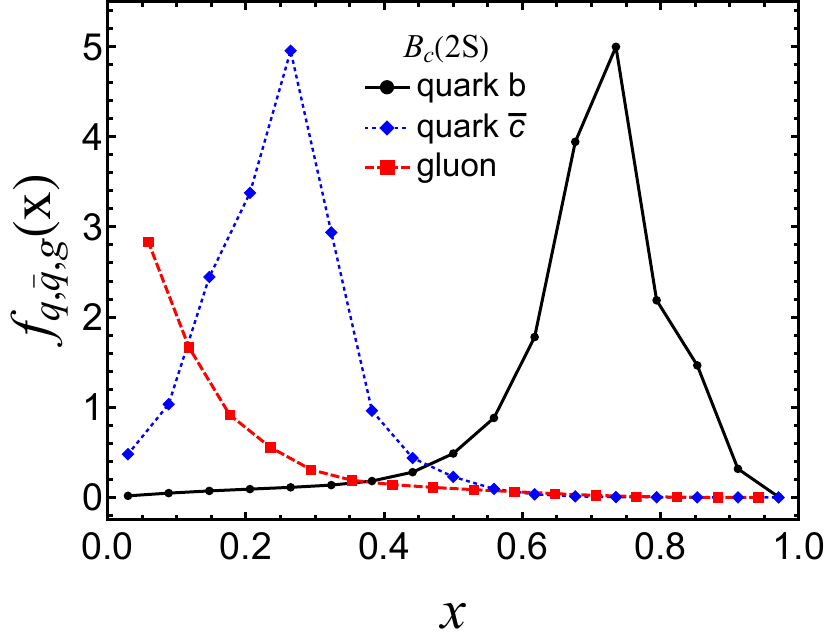}}\quad
		\subfloat[$2^3\mathrm{S_1}$]{\includegraphics[width=0.36\linewidth]{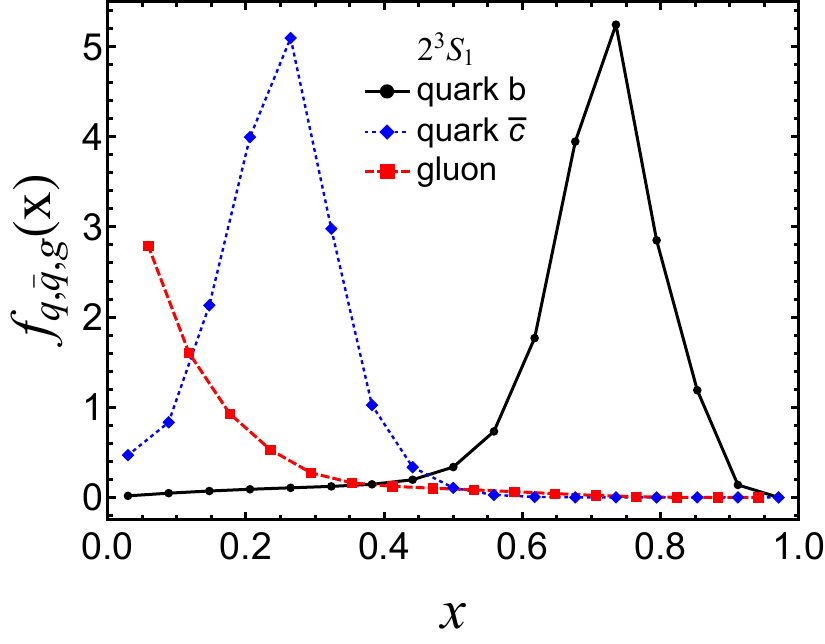}}
		\caption{The unpolarized PDFs at model scale for the heavy mesons $M_J=0$ include contributions from both $|q\bar{q}\rangle$ and $|q\bar{q}g\rangle$ Fock sectors, see text. Panels (a)-(h): charmonium states; panels (i)-(p): bottomonium states; and panels (q)-(x): $\rm B_c$ states. The solid black lines and dashed red lines represent the PDFs of quark and gluon, respectively. For $\rm B_c$ states, the solid black lines, dotted blue lines, and dashed red lines represent the PDFs of $b$ quark, $\bar{c}$ quark, and gluon, respectively.}
		\label{2pdf}
	\end{figure*}

	\subsection{Parton Distribution Function}
	As distinct from the PDAs, the parton distribution functions (PDFs) encode the structural information of hadrons on the probability level. They provide the nonperturbative input for perturbative calculations of the inclusive processes, most notably the Deep Inelastic Scattering (DIS)~\cite{PhysRevD.22.2157}. Physically, the PDF $f_i(x)$ represents the probability density of finding a parton of flavor $i$ carrying a longitudinal momentum fraction $x$ of the parent hadron.
	Although spin-dependent PDFs exist for systems with total angular momentum $J \ge 1$, in this work we focus on the unpolarized distributions. For simplicity, for $J\ge1$ states, we use the $M_J=0$ component to calculate their unpolarized PDFs. The PDFs depend on the energy scale on which the hadron is probed in the inclusive processes.
	At our model scale, the quark PDFs can be expressed in terms of the LFWFs of the $|q\bar{q}\rangle$ and $|q\bar{q}g\rangle$ Fock sectors as~\cite{PhysRevD.102.014020,doi:10.1142/9789811219313_0099,Lan:2021wok},
	\begin{equation}
		f_{q(\bar{q})}\left(x\right)= \sum_{\mathcal{N}, s_i}\int\left[d^{3} p\right]_\mathcal{N} \delta(x-x_1) |\Psi^{\{s_i\}}_{\mathcal{N}}\left(\{x_i,\vec{p}_{\perp i}\}\right)|^2.
	\end{equation}
	
	Similarly, the gluon PDF can be calculated from the LFWF in the $|q\bar{q}g\rangle$ Fock sector,
	\begin{equation}
		f_{g}(x)=\sum_{s_i} \int\left[d^{3} p\right]_3\delta(x-x_3)|\Psi^{\{s_i\}}_{3}\left(\{x_i,\vec{p}_{\perp i}\}\right)|^2.
	\end{equation}
	
	Our PDFs satisfy the following sum rules regarding particle number and momentum conservation,
	\begin{equation}
		\begin{aligned}
			&\int_0^1 f_i(x) dx = N_i,\\
			\sum_{i=q,\bar{q},g} &\int_0^1 x f_i(x) dx = 1,
		\end{aligned}
	\end{equation}
	with $N_i$ being the number of quarks of flavor $i$ in the heavy meson.
	
	In this work, we take the UV regulator $\Lambda_{\mathrm{UV}}\simeq b \sqrt{N_{\max }}$ of the 2D-HO basis as an approximate estimate of our model scale~\cite{y.li1}. At $N_{\max}=12$, the model scales ($\mu_0$) for the charmonium, bottomonium, and $\rm B_c$ meson PDFs are $\mu_0^{c\bar{c}}=4.257$ GeV, $\mu_0^{b\bar{b}}=6.357$ GeV, and $\mu_0^{b\bar{c}}=5.411$ GeV, respectively. The resulting PDFs incorporate contributions from both the leading $|q\bar{q}\rangle$ and the next-to-leading $|q\bar{q}g\rangle$ Fock sectors. The presence of the $|q\bar{q}g\rangle$ Fock sector leads to a nontrivial distribution of the dynamical gluon at the model scale. Compared to previous BLFQ calculations with only the valence sector in the basis~\cite{y.li1,s.tang,PhysRevD.102.014020}, our model scale is expected to be higher. By explicitly including the dynamical gluon degrees of freedom at the model scale, our calculation potentially offers a more realistic description of both the quark and gluon PDFs at evolved scales.
	
	In Fig.~\ref{2pdf} we compare the quark and gluon PDFs for the charmonium, bottomonium, and $\rm B_c$ states at the model scale. For S-wave charmonium and bottomonium states, the quark PDFs are peaked at $x=0.5$ and show a slight enhancement in the small-$x$ region due to the contribution from the $|q\bar{q}g\rangle$ Fock sector. The presence of the gluon breaks the symmetry of the valence quark distribution with respect to $x=0.5$ since the gluon carries a finite fraction of the total longitudinal momentum. Similar to the PDAs, the quark PDFs become narrower as the system mass increases from charmonium to bottomonium states. For $\rm B_c$ mesons, the $b$-quark PDFs are peaked at $x > 0.5$ due to the $b$-quark's heavier mass than the $c$ quark. Therefore, it carries a larger fraction of the longitudinal momentum.
	For the radially excited states, the quark PDFs inherit the nodal structure of the radial excitations from the LFWFs.
	Taking the charmonium system as an example, we observe a distinct double-peak structure for the $\chi_{c0}(1P)$ and $h_c(1P)$ states, in contrast to the single-peak structure of the $\chi_{c1}(1P)$ and $\chi_{c2}(1P)$. This difference arises from their different excitation modes: The $\chi_{c0}(1P)$ and $h_c(1P)$ are excited in the longitudinal direction, cf.~panels (c)-(f) in Fig.~\ref{wave}, leading to the double-peak structure in the longitudinal momentum distribution, whereas the $\chi_{c1}$ and $\chi_{c2}$, being excited in the transverse direction, do not exhibit such double-peak structures in their PDFs.
	
	The gluon PDFs exhibit the characteristic peak structure in the small-$x$ region due to their smaller (effective) mass compared to the heavy quarks and decrease rapidly as $x$ increases. The overall $x$-dependence is qualitatively similar to that found in the light meson~\cite{Lan:2025fia} and nucleon~\cite{PhysRevD.108.094002} systems. 

	Within the same meson system, the gluon distributions for the excited states are notably enhanced compared to the ground state, which suggests that as the system becomes less tightly bound in excited states, the probability of gluon emission increases. This trend is particularly evident in the small-$x$ region, where the excited states exhibit a more pronounced peak.
	Comparing different heavy meson systems, we observe a systematic suppression of the gluon distribution as the constituent quark mass increases. From charmonium to bottomonium, the amount of gluon content decreases, and the gluon PDFs decay more rapidly with increasing $x$. This behavior can be explained from the fact that the gluon radiation is suppressed by the heavy quark mass~\cite{Yu_L_Dokshitzer_1991}. As the mass of the heavy quark increases, the relative weight of the $|q\bar{q}g\rangle$ components is suppressed, leading to a dominance of the valence $|q\bar{q}\rangle$ sector.
	This suppression of the gluon content in heavier systems is consistent with the probability of the $|q\bar{q}\rangle$ sector, $N_{q\bar{q}}^2$, presented in Table~\ref{norm}, where we observed an increasing trend with respect to the heavy quark mass and the excitation energy. In the excited states of the $B_c$ system, however, the gluon PDFs peak more prominently in the small-$x$ region compared to both the corresponding charmonium and bottomonium states, which suggests that the cancellation between gluon radiation from the quark and antiquark is suppressed in this asymmetric system.

	The average longitudinal momentum fraction, denoted as $\langle x \rangle$, characterizes how the total momentum of the meson is partitioned among its constituents. It is defined as the first moment of the unpolarized PDF,
	\begin{equation}
		\langle x_i \rangle = \int_0^1 x f_i(x) dx,
	\end{equation}
	where the subscript $i$ refers to the specific parton species (quark, antiquark, or gluon).
	
	With the unpolarized PDFs contributed from both Fock sectors, we calculate the average longitudinal momentum fraction carried by the quark, antiquark, and gluon in the heavy meson systems. The results are summarized in Table~\ref{component}.
	
	\begin{table}[H]
		\centering
		\caption{The average longitudinal momentum fraction carried by the quarks and gluon as well as the probability of the $|q\bar{q}g\rangle$ Fock sector, $1-N_{q\bar{q}}^2$, in heavy mesons.}
		\begin{tabular}{c*{4}{>{\centering\arraybackslash}p{5em}}}
			\toprule
			\ &$1-N_{q\bar{q}}^2$&$\langle x_q \rangle$&$\langle x_{\bar{q}} \rangle$&$\langle x_{g} \rangle$\\
			\midrule
			$\eta_c(\mathrm{1S})$	&0.357&0.454&0.454&0.091\\
			$J/\psi(\mathrm{1S})$	&0.361&0.456&0.456&0.088\\
			$\chi_{c0}(\mathrm{1P})$&0.467&0.443&0.443&0.114\\
			$\chi_{c1}(\mathrm{1P})$&0.499&0.447&0.447&0.106\\
			$h_c(\mathrm{1P})$		&0.469&0.448&0.448&0.103\\
			$\chi_{c2}(\mathrm{1P})$&0.497&0.447&0.447&0.107\\
			$\eta_c(\mathrm{2S})$	&0.496&0.449&0.449&0.102\\
			$\psi(\mathrm{2S})$		&0.471&0.451&0.451&0.097\\
			\midrule                                  
			$\eta_b(\mathrm{1S})$	&0.189&0.481&0.481&0.038\\
			$\Upsilon(\mathrm{1S})$	&0.185&0.481&0.481&0.037\\
			$\chi_{b0}(\mathrm{1P})$&0.287&0.475&0.475&0.049\\
			$\chi_{b1}(\mathrm{1P})$&0.310&0.476&0.476&0.048\\
			$h_b(\mathrm{1P})$		&0.289&0.477&0.477&0.047\\
			$\chi_{b2}(\mathrm{1P})$&0.310&0.476&0.476&0.048\\
			$\eta_b(\mathrm{2S})$	&0.312&0.477&0.477&0.045\\
			$\Upsilon(\mathrm{2S})$	&0.302&0.479&0.479&0.043\\
			\midrule                                  
			$\rm B_c(\mathrm{1S})$	&0.276&0.697&0.248&0.055\\
			$1^3\mathrm{S_1}$		&0.278&0.702&0.243&0.054\\
			$1^3\mathrm{P_0}$		&0.371&0.686&0.247&0.066\\
			$1^3\mathrm{P_1}$		&0.371&0.708&0.228&0.063\\
			$1^1\mathrm{P_1}$		&0.395&0.691&0.246&0.064\\
			$1^3\mathrm{P_2}$		&0.395&0.708&0.228&0.064\\
			$\rm B_c(\mathrm{2S})$	&0.409&0.692&0.246&0.062\\
			$2^3\mathrm{S_1}$		&0.396&0.697&0.244&0.059\\
			\bottomrule
			\label{component}
		\end{tabular}
	\end{table}
	Compared to quarks, the gluons carry a small fraction of the total longitudinal momentum despite their sizable presence at the model scale. We note that within the same heavy meson system, the gluon contribution $\langle x_g \rangle$ does not increase monotonically with the excitation energy: The gluon carries a larger fraction of the longitudinal momentum in the orbitally excited $\mathrm{1P}$ states compared to both the ground states and the radially excited $\mathrm{2S}$ states. Compared to the $\mathrm{1S}$ states, a larger proportion of the longitudinal momentum fraction is carried by the gluon in the $\mathrm{2S}$ states. Furthermore, the gluon carries a smaller longitudinal momentum fraction in the vector states compared to their pseudoscalar partners for both $\mathrm{1S}$ and $\mathrm{2S}$ states.
	
	Comparing different heavy quarkonium systems, we see the average quark (gluon) momentum fraction increases (decreases) as the heavy quark mass increases. This is again consistent with the picture that the heavier systems are closer to the non-relativistic limit, where the valence quarks carry the majority of the momentum.

	\section{SUMMARY}\label{sec4}

	In this work, we present the first application of the Basis Light-Front Quantization (BLFQ) approach to the heavy quarkonium and $\rm B_c$ meson systems with one dynamical gluon included in the basis. Our effective Hamiltonian features a confining potential in the leading Fock sector, including both transverse and longitudinal components, motivated by the soft-wall AdS/QCD. In addition, in the leading Fock sector, we include the quark-gluon vertex and the instantaneous gluon interaction from the light-front QCD Hamiltonian. We also implement a Fock-sector dependent renormalization scheme to account for the quark self-energy interaction induced by the dynamical gluon.
	
	By solving the eigenvalue problem of the input light-front Hamiltonian, we obtain the mass spectra and light-front wave functions (LFWFs) of the heavy mesons. We determine the input parameters by fitting the mass spectra of low-lying charmonium and bottomonium states. The input parameters for the $\rm B_c$ system are interpolated from those for the charmonium and bottomonium systems. We find a reasonable agreement between the resulting mass spectrum for the low-lying states with the experimental data. In addition, we find that the probability to find a dynamical gluon in the heavy mesons at this level of basis truncation is rather substantial at around 40-50$\%$ (20-30)$\%$ for the charmonium (bottomonium) states.
	
	Utilizing the obtained LFWFs, we compute a large set of observables characterizing the transverse and longitudinal structure of the heavy mesons, including electromagnetic form factors, charge radii, decay constants, Parton Distribution Amplitudes (PDAs), and unpolarized Parton Distribution Functions (PDFs). Our results are qualitatively consistent with those from lattice QCD, Dyson-Schwinger Equation (DSE), and earlier BLFQ calculations based on a one-gluon-exchange effective interaction. In addition, we provide predictions for the gluon PDFs at the model scale. Our results suggest that the gluons are mostly distributed in the small-$x$ region, which is qualitatively similar to the gluon distribution found in the light meson and nucleon systems. Compared to the previous BLFQ calculations based on the one-gluon exchange effective interaction, the explicit inclusion of the dynamical gluon in the nonperturbative calculation represents a major step towards describing the structure of heavy mesons from QCD first principles. Our results confirm the feasibility of the light-front Hamiltonian approach to the heavy meson systems beyond the valence Fock sector and motivate its application to other hadronic systems.
	
	Based on the current work, systematic improvements can be conducted in two directions: First, we plan to include higher Fock sectors containing sea quarks and multiple gluons, such as $|q\bar{q}q\bar{q}\rangle$ and $|q\bar{q}gg\rangle$, and the associated QCD interactions in our calculation. This extension will not only offer a more complete picture of the quark and gluon distribution in the heavy mesons but also open up the possibility of investigating higher excited states and exotic hadron states. Furthermore, since the LFWFs encode the full information of the hadron state, we aim to extend our analysis to the three-dimensional structure of heavy mesons and study the observables such as the generalized parton distributions (GPDs) and the transverse-momentum-dependent parton distributions (TMDs). Through these observables, the light-front Hamiltonian approach will hopefully be able to provide a comprehensive tomography of heavy mesons in connection with QCD first principles.
	
	\section{ACKNOWLEDGMENTS}\label{sec5}
	We thank Siqi Xu, Jiangshan Lan, Yang Li, and Chandan Mondal for valuable discussions. X. Z. is supported by National Natural Science Foundation of China, Grant No.~12375143, by new faculty startup funding by the Institute of Modern Physics, Chinese Academy of Sciences, by Key Research Program of Frontier Sciences, Chinese Academy of Sciences, Grant No.~ZDB-SLY-7020, by the Natural Science Foundation of Gansu Province, China, Grant No.~20JR10RA067, by International Partnership Program of the Chinese Academy of Sciences, Grant No.~016GJHZ2022103FN, by National Key R$\&$D Program of China, Grant No.~2023YFA1606903 and by the Strategic Priority Research Program of the Chinese Academy of Sciences, Grant No.~XDB34000000. J. P. V. is supported in part by the Department of Energy under Grant No.~DE-SC0023692. A portion of the computational resources was also provided by the Gansu Computing Center.
	\bibliography{apstemplate}

\end{document}